\pdfoutput=1
\RequirePackage{ifpdf}
\ifpdf 
\documentclass[pdftex]{sigma}
\else
\documentclass{sigma}
\fi

\numberwithin{equation}{section}

\newcommand{\p}{\partial}
\def\sq2{\sqrt{2}}

\def\sl2{{\rm sl}(2, {\mathbb C})}
\def\SLN{{\rm SL}(N, {\mathbb C})}
\def\SLT{{\rm SL}(2, {\mathbb C})}

\def\mC{{\mathbb C}}
\def\mZ{{\mathbb Z}}

\def\frak{\mathfrak}
\def\gC{{\frak C}}
\def\gg{{\frak g}}

\def\gb{{\frak b}}
\def\gL{{\frak L}}
\def\gk{{\frak k}}

\def\gh{{\frak h}}
\def\gH{{\frak H}}
\def\gt{{\frak t}}
\def\gn{{\frak n}}
\def\gM{{\frak M}}

\def\baal{\bar{\alpha}}
\def\babe{\bar{\beta}}

\def\bfe{{\bf e}}

\def\bfu{{\bf u}}

\def\bfs{{\bf s}}

\def\bfx{{\bf x}}

\def\clA{\mathcal{A}}
\def\clC{\mathcal{C}}

\def\clE{\mathcal{E}}

\def\clU{\mathcal{U}}
\def\clT{\mathcal{T}}
\def\clO{\mathcal{O}}
\def\clH{\mathcal{H}}
\def\clK{\mathcal{K}}

\def\clL{\mathcal{L}}
\def\clM{\mathcal{M}}

\def\clO{\mathcal{O}}
\def\clP{\mathcal{P}}
\def\clQ{\mathcal{Q}}
\def\clS{\mathcal{S}}
\def\clV{\mathcal{V}}
\def\clW{\mathcal{W}}
\def\clZ{\mathcal{Z}}
\newcommand{\ran}{\rangle}
\newcommand{\lan}{\langle}

\begin{document}

\allowdisplaybreaks

\renewcommand{\PaperNumber}{095}

\FirstPageHeading

\ShortArticleName{Hecke Transformations of Conformal Blocks in WZW Theory.~I}

\ArticleName{Hecke Transformations of Conformal Blocks in WZW Theory.~I.~KZB Equations for Non-Trivial Bundles}

\Author{Andrey M.~LEVIN~$^{\dag \ddag}$, Mikhail A.~OLSHANETSKY~$^{\ddag}$,
Andrey V. SMIRNOV~$^{\ddag\S}$\\ and Andrei V.~ZOTOV~$^{\ddag}$}

\AuthorNameForHeading{A.M.~Levin, M.A.~Olshanetsky, A.V.~Smirnov and A.V.~Zotov}

\Address{$^{\dag}$~Laboratory of Algebraic Geometry, GU-HSE, 7 Vavilova Str., Moscow, 117312, Russia}
\EmailD{\href{mailto:alevin57@gmail.com}{alevin57@gmail.com}}

\Address{$^{\ddag}$~Institute of Theoretical and Experimental Physics, Moscow, 117218, Russia}
\EmailD{\href{mailto:olshanet@itep.ru}{olshanet@itep.ru},
\href{mailto:asmirnov@itep.ru}{asmirnov@itep.ru},
\href{mailto:zotov@itep.ru}{zotov@itep.ru}}

\Address{$^\S$~Department of Mathematics, Columbia University, New York, NY 10027, USA}

\ArticleDates{Received July 14, 2012, in f\/inal form November 29, 2012; Published online December 10, 2012}

\Abstract{We describe new families of the Knizhnik--Zamolodchikov--Bernard (KZB) equations related to the WZW-theory
corresponding to the adjoint $G$-bundles of dif\/ferent topological types
over complex curves $\Sigma_{g,n}$ of genus $g$
with $n$ marked points.
The bundles are def\/ined by their characteristic classes~-- elements of
$H^2(\Sigma_{g,n},\mathcal{Z}(G))$, where $\mathcal{Z}(G)$ is a center of the simple complex Lie group~$G$.
The KZB equations are the horizontality condition for the projectively f\/lat connection (the KZB connection)
def\/ined on the bundle of conformal blocks over the moduli space of curves.
The space of conformal blocks has been known to be decomposed
into a few sectors corresponding to the characteristic classes of the underlying bundles.
The KZB connection preserves these
sectors.
In this paper we construct the connection explicitly for elliptic curves with marked points and
prove its f\/latness.}


\Classification{14H70; 32G34; 14H60}

\section{Introduction}

The Knizhnik--Zamolodchikov--Bernard (KZB) equations \cite{Be2,Be1, KZ} are a system
of dif\/ferential equations for conformal
blocks in a conformal f\/ield theory.
Here we consider the WZW theory of the level $k$, related to a simple complex Lie group~$G$
and def\/ined on a Riemann surface $\Sigma_{g,n}$ of genus $g$ with $n$ marked points $(z_1,z_2,\ldots,z_n)$.
To describe this
model, one should def\/ine a~$G$-bundle over $\Sigma_{g,n}$.
Topologically, the $G$-bundles are def\/ined by their characteristic
classes.
Let~$\clZ(G)$ be a~center of $G$ and $G^{\rm ad}=G/\clZ(G)$.
The characteristic classes are obstructions to lift the
$G^{\rm ad}$-bundles to the $G$-bundles.
They are elements of the cohomology group $H^2(\Sigma_g,\clZ(G))\sim\clZ(G)$
\cite{LOSZ}\footnote{See (4.1)--(4.3) in~\cite{LOSZ}.}.
If $\bar{G}$ is the corresponding simply connected group (the universal
covering with the natural group structure) and $G=\bar{G}/\clZ^\vee(G)$,
then elements from $H^2(\Sigma_g,\clZ^\vee(G))$ are
obstruction to lift the $G$-bundles to the $\bar{G}$-bundles.
In particular, consider $G=\operatorname{Spin}(N)$ and
${\rm SO}(N)=\operatorname{Spin}(N)/\mZ_2$.
Then $H^2(\Sigma_g,\mZ_2)\sim\mZ_2$ def\/ines the Stiefel--Whitney classes
of the ${\rm SO}(N)$-bundles over~$\Sigma_g$.

For generic bundles the WZW theories were studied in \cite{FGK,Ho}.
The aim of this paper is to def\/ine the KZB equations in
these theories.
The KZB equations have a large range of applications in mathematics.
In particular, on the critical level they produce
Hamiltonians of the quantum Hitchin system \cite{Ne11,Hi1,Ne13,Ne12},
while in the classical limit they lead to the
monodromy-preserving equations \cite{Ha, Ta1,LO,Re,Ta2}.
In this way, we obtain new classes of these systems.

The KZB equations are described in the following way.
Consider the highest weight representations $V_{\mu_a}$ (${\mu_a}$ are
the highest weights) of $G$ attached to the marked points.
For a positive integer $k$ def\/ine the integrable module $\hat
V_{\mu_a}$ of level $k$ of the centrally extended loop group $D^\times\to G$,
where $D^\times=D\setminus  z_a$ is a punctured
disk around the marked point $z_a$.
The conformal blocks are linear functionals $\hat V^{[n]}\equiv\hat
V_{\mu_1}\otimes\dots\otimes\hat V_{\mu_n}\to\mC$ satisfying some additional conditions (the Ward identities).
Let
$\clC_G(\hat V^{[n]})$ be a space of conformal blocks.
This space depends on parameters~-- the complex structure of
$\Sigma_{g,n}$, and in this way forms a bundle over the moduli space $\gM_{g,n}$ of complex structures.
There exists a
projectively f\/lat connection in this bundle (the KZB connection).
Then the meaning of the KZB equations is that the
conformal blocks are the horizontal sections of the KZB connection.
The KZB equations were derived originally for the genus
zero case by Knizhnik and Zamolodchikov~\cite{KZ}
and were generalized later to arbitrary genus by Bernard~\cite{Be2, Be1}.
In subsequent years the KZB equations was studied in a number of works \cite{APW,ER,Fe,Fuchs,Hi,I}.

If the cocenter $\clZ^\vee(G)=\operatorname{Ker}\bar{G}\to G$
is non-trivial then the integrable module is a sum of sectors,
corresponding to the characteristic classes of the underlying bundles
\[
\hat V_\mu=\bigoplus_{j=0}^{N-1}V_\mu^{(j)},\qquad N=\operatorname{ord}\clZ(G).
\]
In terms of the spectra the WZW theory this was studied essentially in~\cite{FGK}.
Similarly, the conformal blocks are also a sum of dif\/ferent sectors.
In each sector one can def\/ine
the KZB connection.

\emph{The aim of this paper} is to construct explicitly the
KZB connections in all sectors of conformal blocks for the WZW theory def\/ined on elliptic curves.
The compatibility conditions (horizontality of the KZB connection) are verif\/ied explicitly.

The KZB connection in the trivial sector was studied in \cite{FW}.
This construction is based on the classical dynamical
$r$-matrix with the spectral
parameter living on the elliptic curve.
The $r$-matrices of this type related
to the trivial sector were classif\/ied by Etingof and Varchenko~\cite{EV}.
Recently, we have classif\/ied the dynamical elliptic $r$-matrices as sections of some bundles of an arbitrary
topological type over elliptic curves~\cite{LOSZ}.
It turned out that the dynamical parameters
of the $r$-matrices are elements of the moduli spaces of the bundles.
It allows us to def\/ine the KZB connection in these cases.

Dif\/ferent approach to classif\/ication of elliptic
$r$-matrices was proposed in \cite{ES1,ES2,Feher}
and the corresponding
KZB connection was also constructed in \cite{ES1,ES2}.
The staring point of last approach is an automorphism
of the extended Dynkin diagram.
In our construction we considered only those automorphisms
that isomorphic to elements of the center $\clZ$.
In this case we come to the same $r$-matrices
and the KZB equations as in \cite{ES1,ES2}.
For~$A_n$,~$D_n$ and~$E_6$ algebras there exists another type automorphisms.
So far the underlying vector bundle structure is
unclear.
It should be noted that in \cite{ES1,ES2} the derivation of the KZB equation is based on the representations
of conformal blocks as twisted traces of intertwiners.
We will come to this representation
in the forthcoming paper where the Hecke transformation of conformal blocks will be considered (see below).

For the $\SLN$ WZW model on elliptic curves the KZB equation in the similar to our form was described in \cite{KT}.
The
authors considered a particular type of bundles that lead to the Belavin--Drinfeld classical $r$-matrix.
In this case the
corresponding KZB equation has not dynamical parameter and similar to the KZ equation.
However, if $N$ is not a prime number
there exist $r$-matrices and the corresponding KZB equations intermediate between Felder and Belavin--Drinfeld cases.

In the subsequent paper we will describe the transformation operators that intertwine the dif\/ferent sectors
(the Hecke transformations).
It is worthwhile to notice that in the classical case these transformations provide a passage from the
elliptic Calogero--Moser system to integrable Euler--Arnold top \cite{LOZ_1,LOZ_3,int13} (see also
\cite{int2,int12,int13, int11}).
For arbitrary characteristic classes these type of models were described in \cite{LOSZ2}.
Dif\/ferent aspects and applications of the Hecke transformations to integrable systems and related topics (such as
Painlev\'e--Schlesinger equations \cite{iso24,iso21,iso12,iso22,iso23,iso41,iso42, iso11}, monopoles
\cite{KW22,KW12,KW13,KW23, KW11,KW21,KW14}, quadratic Poisson structures \cite{dirac1,dirac2}, applications to AGT
conjecture \cite{agt3, agt1,agt2} etc.) can be found in wide range of literature.

The paper has the following structure.
In Section~\ref{section2} we consider a general setting
of the KZB equations related to arbitrary curves $\Sigma_{g,n}$
and arbitrary characteristic class of the bundles.
In Section~\ref{section3} the space of conformal blocks is described.
In Section~\ref{section4} we consider the genus one case
in detail.
The proofs of main relations (Propositions~\ref{predl1} and~\ref{predl2})
and information about the special
basis in simple Lie algebras as well as the elliptic functions identities are given in the appendices.

\section{Loop algebras, loop groups and integrable modules}\label{section2}

\subsection{Loop algebras and loop groups}\label{section2.1}

Let $\bar{G}$ be a simply-connected simple complex Lie group and $\clZ=\clZ\big(\bar{G}\big)$ is the center of $\bar{G}$.
For all
simply-connected groups
($\SLN$, ${\rm Sp}_N$, $E_6$, $E_7$ and $\operatorname{Spin}_N$ except $N=4n$), the center is a cyclic group.
For $\operatorname{Spin}_{4N}\clZ=\mZ_2\oplus\mZ_2$.
The adjoint group is the quotient group $G^{\rm ad}=\bar{G}/\clZ$.
Assume for simplicity that $\clZ$ is a cyclic group $\mZ_l$ of order $l$.

Let $K$ be a maximal compact subgroup of $\bar{G}$ and $T$ is the Cartan torus of $K$.
Consider the homomorphisms of $S^1\to T$
\[
\bfe(\varphi)\to\big(\bfe(\gamma_1\varphi),\bfe(\gamma_2\varphi),\ldots,\bfe(\gamma_l\varphi)\big)\in T,
\qquad
\bfe(\varphi)=\exp(2\pi\imath\varphi).
\]
$P^\vee=\{\gamma=(\gamma_1,\ldots,\gamma_l)\}$ is a coweight lattice
in the Cartan subalgebra $\gh^{K}=\operatorname{Lie}(T)$ and in
$\gh\subset\gg=\operatorname{Lie}\big(\bar{G}\big)$.
Let $Q^\vee$ be the coroot lattice $(Q^\vee\subseteq P^\vee)$.
The center $\clZ\big(\bar{G}\big)$ is
isomorphic to the quotient group $\clZ\sim P^\vee/Q^\vee$.
In particular,
if $\varpi^\vee\in P^\vee$ is a coweight such that $l\varpi^\vee\in Q^\vee $,
then the $\clZ\sim\mZ_l$.
It is generated by the element
$\bfe(\varpi^\vee)=\exp(2\pi\imath\varpi^\vee)\in T$.
For $\operatorname{Spin}_{4N}$ the center is
generated by two coweights, corresponding to the left and right spinor representations.

Let $\gh$ be a Cartan subalgebra of $\gg$ and $\{\alpha\}=R\in\gh^* $
is the root system \cite{Bou}. There is
the root decomposition of $\gg$,
\[
\gg=\gh\oplus\sum_{\alpha\in R}\gg^\alpha,
\qquad
\operatorname{ad}_X\gg^\alpha=\lan X,\alpha\ran\gg^\alpha,
\qquad
X\in\gh.
\]
$R$ is an union of positive and
negative roots $R=R_+\cup R_-$ with respect to some ordering in~$\gh^*$.
Let $\Pi=\{\alpha_1,\ldots,\alpha_l\}$ be a basis
of simple roots in~$R$.
The dual system $\Pi^\vee=\{\alpha^\vee_1,\ldots,\alpha^\vee_l\}$
($\lan\alpha_j,\alpha_k^\vee\ran=\delta_{jk}$) forms a basis in~$\gh$.

Let $t$ be a coordinate in $\mC$.
Def\/ine the loop group $L(G)=\{\mC^*\to G\}=\{g(t)\}$ such that $g(t)$ has a f\/inite
order poles when $t\to 0$.
In other words, $L(G)$ is the group of Laurent polynomials $L(G)=G\otimes\mC[[t,t^{-1}]$.
There is a central extension $\hat L(\gg)$ of $L(\gg)=\gg\otimes\mC[[t,t^{-1}]$
\begin{gather}\label{cea}
\hat L(\gg)=L(\gg)\oplus \mC K,
\end{gather}
def\/ined by a two-cocycle $c(X\otimes f,Y\otimes g)=(X,Y)\operatorname{Res}(gf'dt)$.

The set of the af\/f\/ine roots if of the form: $R^{\rm af\/f}=\{\hat\alpha=\alpha+n$, $n\in\mZ$, $n\neq 0\}$.
Let $\{\gh_\alpha\}$ be the
basis of simple coroots in $\gh$.
Then the analog of the root decomposition for the loop algebra has the following form
\[
L(\gg)=\gg+\sum_{n\neq 0}\sum_{\alpha\in\Pi}\gh_{\alpha}t^n+ \sum_{\tilde\alpha\in R^{\rm af\/f}} \gg^{\hat\alpha},
\qquad
 \gg^{\hat\alpha}=x_\alpha e_{\hat\alpha}=x_\alpha e_\alpha t^n .
\]
Let $-\alpha_0$ be the highest root $-\alpha_0\in R_+$.
The system of simple af\/f\/ine roots is $\hat\Pi=\Pi\cup(-\alpha_0+1)$. It is a
basis in $R^{\rm af\/f}$.
Consider the positive loop subalgebra
\begin{gather}\label{lgp}
L^+(\gg)=\big(\gb+\gg \otimes t\mC[[t]]\big),
\end{gather}
where $\gb=\gh\oplus\sum\limits_{\alpha\in R^+}\gg^\alpha$ is the positive Borel subalgebra.
Let also
\begin{gather}\label{lgm}
L^-(\gg)=\big(\gn_-+\gg\otimes t^{-1}\mC[t^{-1}]\big),
\qquad
\gn_-= \sum_{\alpha\in R^-}\gg^\alpha.
\end{gather}
Then $\hat L(\gg)$ (\ref{cea}) is the direct sum
\begin{gather}\label{hl}
\hat L(\gg)=L^-(\gg)\oplus L^+(\gg)\oplus\mC K.
\end{gather}
Each summand is a Lie subalgebra of $\hat L(\gg)$.
There are two types of the af\/f\/ine Weyl groups: $W_P=W\ltimes P^\vee$ and $W_Q=W\ltimes Q^\vee$,
where $W$ is the Weyl group of $\gg$,
\begin{gather}\label{awg}
W_P=\left\{\hat w=wt^\gamma,\;w\in W,\;\gamma\in P^\vee\right\},
\qquad
W_Q=\left\{\hat w=wt^\gamma,\;w\in W,\;\gamma\in Q^\vee\right\}.
\end{gather}
They act on the root vectors as
$ e_{\hat\alpha}= e_\alpha t^n\to e_{\hat w(\hat\alpha)}=e_{w(\alpha)}t^{n+\lan\gamma,\alpha\ran}$.
The loop groups $L(G)=G\otimes\mC[[t,t^{-1}]]$ have the Bruhat decomposition \cite{PS}.
Def\/ine subgroups
\begin{gather}\label{bpl}
L^+(G)=\big\{g_0+g_1t+\cdots\big\},\quad g_j\in G,\;\; g_0=b\in B,\quad\text{is the positive Borel subgroup},
\\
\label{bne}
N^-(G)=\big\{n_-+g_1t^{-1}+\cdots\big\},\quad n_-\in N_-, \quad \text{is~the~negative~nilpotent~subgroup},
\\
\label{pns}
N^+(G)=\big\{n_++g_1t+\cdots\big\},\quad n_+\in N_+, \quad \text{is~the~positive~nilpotent~subgroup}.
\end{gather}
The Bruhat decomposition takes the form
\begin{gather}\label{PS}
L\big(G^{\rm ad}\big) =\bigcup_{\hat w\in W_P}N^-\big(G^{\rm ad}\big)\hat wL^+\big(G^{\rm ad}\big),
\qquad
L\big(\bar{G}\big) =\bigcup_{\hat w\in W_Q}N^-\big(\bar{G}\big)\hat wL^+\big(\bar{G}\big).
\end{gather}
For a loop $g(t)$ in $G^{\rm ad}$ denote by $\bar g$ its lift to a map from $S^1$ to $\bar{G}$.
This map can be
multivalued, after turning
along the circle the value can be multiplied by some element of the center which we call the monodromy:
$g(e^{2\pi\imath}t)=\bfe(\gamma)g(t)$, ($\bfe(x)=e^{2\pi\imath x}$).
If $\gamma\notin Q^\vee$ then $\zeta=\bfe(\gamma)$ is a~non-trivial element of the center $\clZ$ and the map
$g(t)$ is well def\/ined for $G=G^{\rm ad}$, but not for $\bar{G}$.
In this way we have the representations
\begin{gather}\label{rl}
L\big(G^{\rm ad}\big)=\bigcup_{\gamma\in P^\vee}L_\gamma\big(G^{\rm ad}\big),
\qquad
L_\gamma\big(G^{\rm ad}\big)=\big\{g\big(e^{2\pi\imath}t\big)=\bfe(\gamma)g(t)\big\}.
\end{gather}
If $\gamma_1=\gamma_2+\delta$ for any $\delta\in Q^\vee$ then $\gamma_1$ and $\gamma_2$ lead to the same
monodromies.
We say in this case that $L_{\gamma_1}\big(\bar{G}\big)$ and $L_{\gamma_2}\big(\bar{G}\big)$ are equivalent.
Then from (\ref{rl}) we have
\begin{gather*}
L\big(G^{\rm ad}\big)=\bigcup_{\zeta\in\clZ}L_\zeta\big(G^{\rm ad}\big).
\end{gather*}
In particular, if the center $\clZ\sim\mZ_l$ is generated by a fundamental coweight $\varpi^\vee$, then{\samepage
\begin{gather}\label{lgp2}
L\big(G^{\rm ad}\big)=\bigcup_{j=0}^{l-1}L_j\big(G^{\rm ad}\big),
\qquad
L_j\big(G^{\rm ad}\big)=\big\{g\big(e^{2\pi\imath}t\big)=\bfe(j\varpi^\vee )g(t)\big\},
\end{gather}
and $ L_j\big(G^{\rm ad}\big)=\bfe(j\varpi^\vee)\bigl(L\big(\bar{G}\big)/\clZ\bigr)$.}

Consider the quotient $\operatorname{Fl}^{\rm af\/f}=L\big(G^{\rm ad}\big)/L^+\big(G^{\rm ad}\big)$ \cite{PS}.
It is called the af\/f\/ine f\/lag variety.
Let $\Sigma_{\hat w}$
be an $N^-\big(G^{\rm ad}\big)$-orbit of $\hat w$ in $\operatorname{Fl}^{\rm af\/f}$.
This orbit is dipheomorphic to the intersection
$N^-\big(G^{\rm ad}\big)_{\hat w}=N^-\big(G^{\rm ad}\big)\cap\hat wN^-\big(G^{\rm ad}\big)\hat w^{-1}$.
Therefore, its codimension in $Fl^{\rm af\/f}$ is the length
$l(\hat w)$ of $\hat w$.
It is the number of negative af\/f\/ine roots which $\hat w$ transforms to positive ones
(Theorem~8.7.2 in~\cite{PS}).
The Bruhat decomposition (\ref{PS}) def\/ines the stratif\/ication of $\operatorname{Fl}^{\rm af\/f}$:
\begin{gather}\label{bc}
\operatorname{Fl}^{\rm af\/f}=L\big(G^{\rm ad}\big)\big/L^+\big(G^{\rm ad}\big)=\bigcup_{\hat w\in W_P}\Sigma_{\hat w}.
\end{gather}

\subsection{Integrable modules}\label{section2.2}
Consider a subset of dominant weights
$P^+=\{\mu\in P|\lan\mu,\alpha^\vee\ran\geq 0\text{ for }\alpha^\vee\in\Pi^\vee\}$.
Each dominate weights def\/ine
a $\gg$-module $V_\mu$.
It contains the highest weight vector (HWV) $v_\mu$
such that
\[
Xv_\mu=\lan X,\mu\ran v_\mu\quad  \text{for}\quad X\in\gh,
\qquad
\gg^\alpha v_\mu=0\quad  \text{for}\quad \alpha\in R^+.
\]
Def\/ine the Verma module $\clV_\mu$ of $\hat\clL(\gg)$ associated with $V_\mu$
\cite{Ka}.
Let $I_k=\{\mu\in P^+|\lan\mu,\alpha^\vee_0\ran\leq k\}$ be a subset of dominant weights.
Def\/ine the action of $L^+(\gg)$
(\ref{lgp}) on $V_\mu$: $ (\gg\otimes t\mC[[t]])V_\mu=0$, $KV_\mu=k\operatorname{Id}$,
and $\gb$ acts on $V_\mu$ as described above.
Then $\clV_\mu=U(\hat L(\gg))\otimes_{U(L(\gg)^+)}V_\mu$ is induced, where $\mu\in I_k$.
There is the isomorphism
\[
\clV_\mu\sim U(L^-(\gg))\otimes_\mC v_\mu.
\]
Let $E_{\alpha_0}$ be the root subspace in $\gg$ corresponding to $\alpha_0$.
Consider the maximal submodule $\clS_\mu$ of~$\clV_\mu$ generated by the singular vector
\begin{gather}\label{sv}
\left(E_{\alpha_0}\otimes t^{-1}\right)^{k-\lan\mu,\alpha_0\ran+1}v_\mu.
\end{gather}
The irreducible\emph{ integrable module} $\hat{V}_\mu$ is the quotient
\begin{gather}\label{hV}
\hat{V}_\mu=\clV_\mu/\clS_\mu.
\end{gather}
We identify the module $V_\mu$ with a submodule $V_\mu\otimes 1\hookrightarrow\clV_\mu$.
The integrable module $\hat{V}_\mu$
can be characterized in the following way: \emph{the subspace of $\hat{V}_\mu$ annihilated by the positive subalgebra
$\gg\otimes\mC[[t]]$ is isomorphic to the finite-dimensional $\gg$-module $V_\mu$}
\begin{gather*}
V_\mu\sim\big\{v\in\hat{V}_\mu|(\gg\otimes t\mC[[t]])\cdot v=0\big\}.
\end{gather*}
The group $L(G)$ has a central extension $1\to\mC^*\to\widehat{LG}\to LG\to 1$
corresponding to (\ref{cea}).
The integrable module can be described in terms
of $\widehat{LG}$.
The action of $L^+(G)$ on the HWV has the form
\begin{gather}\label{hvl}
L^+(G)v_\mu=\chi_\mu(b)v_\mu,
\qquad
\lambda v_\mu=\bfe(k)v_\mu,
\qquad
\lambda\in\text{the center }\mC^*,                                                    
\end{gather}
where $\chi_\mu(b)$ is the character of the Borel subgroup $B$.
Then $\hat{V}_\mu$ is generated by the action of~$N^-(G)$~(\ref{bne}) on $V_\mu$.

In this way we describe only ``the trivial sector'' of the $L(G)$-module.
Consider the Bruhat representation for $L\big(G^{\rm ad}\big)$
(\ref{PS}), and let $\hat w=t^\gamma$, $\gamma\in P^\vee$.
Def\/ine the Verma modules with the HWV $t^\gamma v_\mu$,
\begin{gather}\label{img1}
\clV_\mu(\gamma)= U(L^-(\gg))\otimes_\mC t^\gamma v_\mu.
\end{gather}
They have the singular vectors
$\left(E_{\alpha_0}\otimes t^{-1}\right)^{k-\lan\mu\alpha_0\ran+1} t^\gamma v_\mu$ (compare with (\ref{sv})).
Let $\clS_{\mu,\gamma}$ be the maximal
submodules generated by these singular vectors.
Consider the quotient spaces
\begin{gather}\label{img2}
\hat V_\mu(\gamma)=\clV_{\mu}(\gamma)/\clS_{\mu,\gamma},
\end{gather}
and def\/ine their direct sum
\begin{gather}\label{trm1}
\hat{\bf{V}}_\mu=\bigoplus_{\gamma\in P^\vee} \hat{V}_{\mu}(\gamma).
\end{gather}
We say that two subspaces $\hat{V}_{\mu}(\gamma_1)$ and $\hat{V}_{\mu}(\gamma_2)$
are equivalent if $\gamma_1=\gamma_2+\delta$, where
$\delta\in Q^\vee$.
This equivalence leads to the decomposition of $\hat{\bf{V}}_\mu$ (as a $L(G^{\rm ad})$-module)
into a sum of $l=\operatorname{ord}(\clZ(\bar{G}))$ sectors,                                             
\begin{gather}\label{img}
\hat{\bf V}_\mu=\bigoplus_{\zeta\in\clZ} \hat{V}_{\mu}(\zeta).
\end{gather}
Notice that
$(t^\gamma v_\mu)$ is not the HWV with respect to $L^+(\gg)$.
However, it was proved
in \cite{FGK} that there exists a unique element $\hat w=\hat w(\gamma)=t^\delta w\in W_Q$
such that $t^\gamma\hat wv_\mu$ is the
HWV.
We demonstrate it below for $L(\SLT)$.
The elements $\hat w$ and $\gamma$ represent the same element $\zeta\in\clZ$.
Then
we def\/ine the Verma module
\begin{gather}\label{um}
\clV_\mu(\zeta)\sim U(L(\gg)^-)\otimes_\mC( t^\gamma\hat w v_\mu).
\end{gather}
The vector $\left(E_{\alpha_0}\otimes t^{-1}\right)^{k-\lan\mu,w\alpha_0\ran+1}(\gamma\hat wv_\mu)$
is singular and corresponds to the
submodule $\clS_{\mu,\gamma}$.
As in (\ref{hV}) we identify the integrable modules $\clV_\mu(\zeta)/\clS_{\mu,\gamma}$ with
$\hat{V}_\mu(\zeta)$ (\ref{img1}).

Let $\hat{V}^*_\mu$ be the dual module.
The Borel--Weil--Bott theorem for the loop group \cite{PS} states that~$\hat{V}^*_\mu$
can be realized as the space of sections of a line bundle $\clL_\mu$ over the af\/f\/ine f\/lag va\-rie\-ty~(\ref{bc}).
The line bundle is determined by the action $L^+(G)\times \mC^*$ on its sections as in~(\ref{hvl}),
\begin{gather}\label{lbg}
\clL_\mu=\left\{(g,\xi)\sim\left(gb,\chi_\mu\left(b^{-1}\right)\xi\right),\; g\in L(G),\; b\in L^+(G)\right\}.
\end{gather}

\section{Conformal blocks and KZB equation in general case}\label{section3}

\subsection[Moduli space of holomorphic $G$-bundles]{Moduli space of holomorphic $\boldsymbol{G}$-bundles}\label{section3.1}

Let $\clP$ be a principle $G$-bundle over a curve $\Sigma_{g,n}$ of genus $g$ with
$n$ marked points
$\vec{z}=(z_1,\ldots,z_n)$, $(n>0)$, $V$ is a $G$-module
and $E_G=\clP\times_GV$ is the associated bundle.
We consider
the set of isomorphism classes of holomorphic $G$-bundles $\clM_{G,g,n}$ over $\Sigma_{g,n}$
with \emph{the quasi-parabolic
structures at the marked points} \cite{Si}.
They are def\/ined in the following way. A $G$-bundle can be trivialized over small
disjoint disks $D=\bigcup\limits_{a=1}^nD_a$ around the marked points and over $\Sigma_{g,n}\setminus  \vec{z}$.
Therefore,
$\clP$ is def\/ined by the transition holomorphic functions
on $D^\times=\bigcup\limits_{a=1}^n(D_a^\times)$ and
$D_a^\times=D_a \setminus  z_a$.
If
$G(X)$ are the holomorphic maps from $X\subset\Sigma_g$ to $G$,
then the isomorphism classes are def\/ined
as the double coset space
\begin{gather}\label{dcc}
\operatorname{Bun}_G= G(\Sigma_{g,n} \setminus  \vec{z})  \setminus  G(D^\times)/G(D)\sim\clM_{G,g,n}.
\end{gather}
Let $t_a$ be a local coordinate in the disks $D_a$.
Then
$G(D)=\prod\limits_{a=1}^nG(D_a)=\prod\limits_{a=1}^nG\otimes\mC[[t_a]]$
and
\begin{gather}\label{ma}
G(D^\times)=\prod_{a=1}^nL_a(G),
\qquad
L_a(G)=G\otimes\mC[[t_a,t_a^{-1}].
\end{gather}

Let us f\/ix $G$-f\/lags at f\/ibers over the marked points.
The quasi-parabolic structure of the $G$-bundle means that $G(D)$ preserves these $G$-f\/lags.
In other words, $G(D_a)=L^+_a(G)$ (\ref{bpl}).
At the level of the Lie algebra $\operatorname{Lie}(G(D))=\bigoplus_{a=1}^nL_a^+(\gg)$
(\ref{lgp}).
We discuss the Lie algebra $\gg_{\rm out}=\operatorname{Lie}(G(\Sigma_{g,n} \setminus  \vec{z}) )$ below.

Consider the one-point case $\vec{z}=z_0$ in (\ref{ma}).
Let $g(t)\in G[[t,t^{-1}]=G(D^\times_{z_0})$ be the transition
function on the punctured disc $D^\times_{z_0}$ with the local coordinate $t$.
This transition function def\/ines a $G$-bundle.
Its Lie algebra
$\operatorname{Lie}(G(D^\times))=\gg\otimes\mC[[t,t^{-1}]$ assumes the form (see (\ref{dcc}))
\begin{gather}\label{lms}
\operatorname{Lie}(G(D^\times))=\gg_{\rm out}\oplus T\operatorname{Bun}_G\oplus L^+( \gg).
\end{gather}
Introduce a new transition matrix $\tilde g(t)=t^\gamma g(t)$, where $\gamma\in P^\vee$
is an element of the coweight lattice.
It def\/ines a new bundle $\tilde E_G$.
The passage from $E_G$ to $\tilde E_G$ is called \emph{the modification}
of the bundle $E_G$ at the point $z_0$.
The modif\/ication amounts to the passage between dif\/ferent sectors of the integrable
module attached at $z_0$ (see (\ref{img1}), (\ref{img2}), (\ref{trm1})).
Since $t^\gamma\in B$, where~$B$ is the Borel subgroup
$(\gb=\operatorname{Lie}(B)\subset L^+( \gg))$~(\ref{lgp}),
we say that modif\/ication is performed in the ``direction'', consistent with the quasi-parabolic structure at
$z_0$.
In general, it can have an arbitrary direction.
It means that $t^\gamma$ may be replaced by $\operatorname{Ad}_f(t^\gamma)$,
where $f\in G$.
As it was mentioned in Section~\ref{section2.2} there is a unique modif\/ication that
preserves the HWV of the integrable module~$\hat V_\mu$ attached at~$z_0$.

To be a $\bar{G}$-bundle over $\Sigma_g$ the transition matrix $g$ should have
a trivial monodromy $g\big(te^{2\pi i}\big)=g(t)$ around $w$.
If $g(t)$ has a trivial monodromy and $\gamma$ belongs to
the coroot sublattice $Q^\vee$, then~$\tilde g(t)$ also has a trivial monodromy.
Otherwise, the monodromy is an element of the
center~$\clZ\big(\bar{G}\big)$.
For example, let $\gamma=j\varpi^\vee$, where $\varpi^\vee$ generate the group $\mZ_l$, i.e.\
$l\varpi^\vee\in Q^\vee$, while $j\varpi^\vee\notin Q^\vee$ for $j\neq0$, $\operatorname{mod}(l)$.
In this case
\begin{gather}\label{mon}
g\big(te^{2\pi i}\big)=\zeta^j g(t),
\qquad
\zeta=\bfe(\varpi^\vee).
\end{gather}
If $j\neq 0$ then $g(t)$ is not a transition matrix for the $\bar{G}$-bundle.
But it can be considered as
a transition matrix for the $G^{\rm ad}$-bundle, since $G^{\rm ad}=G/\clZ$.
In this case
the $G$-bundle is topologically non-trivial and
$\zeta$ represents \emph{the characteristic class} of $E_G$.
The characteristic class is an obstruction to lift $G^{\rm ad}$-bundle to $G$-bundle.
It is represented by an element
$H^2(\Sigma_g,\clZ)$ \cite{LOSZ}.
Let $\tilde g(t)=g_j(t)=t^{j\varpi^\vee}$.
Then the multiplication by $g_j(t)$ provides a passage in (\ref{lgp2}) from the trivial sector
to the non-trivial sectors
\begin{gather*}
g_j(t)\cdot L_0\big(G^{\rm ad}\big)=L_j\big(G^{\rm ad}\big).
\end{gather*}

In general, we have a decomposition of the moduli space (\ref{dcc}) into sectors
\begin{gather}
 \clM_{G^{\rm ad},g,1}=\bigcup_{\gamma\in P^\vee} \clM^{(\gamma)}_{G^{\rm ad},g,1},\nonumber\\
 \clM^{(\gamma)}_{G^{\rm ad},g,1}
=G\big(\Sigma_{g,n}\setminus w\big)\setminus  G^{\rm ad}_\gamma\otimes\mC[[t,t^{-1}]/G\otimes\mC[[t]],\nonumber\\
 G^{\rm ad}_\gamma\otimes\mC[[t,t^{-1}]=G^{\rm ad}\otimes t^\gamma\mC[[t,t^{-1}].\label{grm}
\end{gather}
In particular, for $\Sigma_{0,1}$ ($\mC P^1\sim\mC\cup\infty$) and the marked point $z_1=0$
this representation is related to the Grothendieck description of the vector bundles over $\mC P^1$.
Let $g_-\in G\otimes\mC[z^{-1}]$, $g_+\in G\otimes\mC[z]$.
Then $ g(z)\in L(G)=(\mC^*\to G)=\{g(z)\}$
has the Birkhof\/f decomposition~\cite{PS}
\begin{gather}\label{bd}
g(z)=g_-z^\gamma g_+, \qquad \gamma\in P^\vee.
\end{gather}
It means that any vector bundle $E_G$ over $\mC P^1$
is isomorphic to the direct sum of the line bundles $\oplus_{i=1}^l \clL_{\gamma_i}$, where $\clL_{\gamma_i}$ is
def\/ined by the transition function $z^{\gamma_i}$, $\gamma=(\gamma_1,\ldots,\gamma_l)$.
If $\gamma\notin Q^\vee E_G$ then has a non-trivial characteristic class.
In fact, the bundle with $\gamma\neq 0$ are
unstable.

Two subsets $\clM^{(\gamma_1)}_{G,g,1}$ and $\clM^{(\gamma_2)}_{G,g,1}$ of the moduli space
correspond to the vector bundles
with the same characteristic class if $\gamma_1=\gamma_2+\beta$, $\beta\in Q^\vee$.
Then the topological classif\/ication of the moduli spaces of the vector bundles by their characteristic classes
follows from (\ref{grm})
\begin{gather}
 \clM_{G,g,1}=\bigcup_{\zeta\in\clZ} \clM^{(\zeta)}_{G,g,1},\nonumber\\
 \clM^{(\zeta)}_{G,g,1}
 =G\big(\Sigma_{g,n}\setminus w\big)\setminus G^{\zeta}\otimes\mC[[t,t^{-1}]/G\otimes\mC[[t]],\nonumber\\
 G^{(\zeta)}\otimes\mC[[t,t^{-1}]= G\otimes t^{j\varpi^\vee}\mC[[t,t^{-1}],
\qquad
\zeta=\bfe\big(j\varpi^\vee\big).\label{grm1}
\end{gather}
Similar representation exists for the space $\clM_{G,g,n}$.

\subsection{Moduli of complex structures of curves}\label{section3.2}

Let $\gM_{g}$ be the moduli space of complex structures of compact curves $\Sigma_g$ of genus $g$.
The moduli space $\gM_{g,n}$
of the complex structures of curves with marked points is foliated over $\gM_{g}$ with f\/ibers $\clU\subset\mC^n$
corresponding to the moving marked points.

An inf\/initesimal deformation of the complex structures is represented by the Beltrami $(-1,1)$ dif\/ferential
$\mu(z,\bar{z})=\mu\frac{\p}{dz}\otimes d\bar{z}$ on $\Sigma_{g,n}$.
In this way $\mu$ is $(0,1)$ form on $\Sigma$ taking values in
$T^{(1,0)}(\clM_{g,n})$ and vanishing at the marked points.
The basis in the tangent space $T(\gM_{g,n})$ is represented by
the Dolbeault cohomology group $H^1(\Sigma_g, \Gamma(\Sigma_g\setminus  \vec z)\otimes \bar K)$,
where $\bar K$ is the anti-canonical class.

Let us compare it with the $\rm\check{C}$ech like construction of
$T\gM_{g,n}$ as a double coset space.
As above, consider
small disks $D_a$ around marked points with local coordinates $t_a$.
Let
$\mC[[t_a,t_a^{-1}]\p_{t_a}$, $\mC[[t_a]]\p_{t_a}$ be vector f\/ields on $D_a^\times$
while $D_a$ and $\Gamma_{(\Sigma_g\setminus \vec{z})}$ is a space of vector f\/ields on
$\Sigma_g\setminus \vec{z}$.
The vector f\/ields from the latter space can have poles of f\/inite
orders at the marked points.
Then
\begin{gather}\label{dccu}
T\gM_{g,n}
=\Gamma_{(\Sigma_g\setminus \vec{z})}\!\left\backslash\;\bigoplus_{a=1}^n\mC[[t_a,t_a^{-1}]\p_{t_a}
\right/\bigoplus_{a=1}^n\mC[[t_a]]\p_{t_a}.
\end{gather}
This construction has the following relation to the Dolbeault description.
We establish correspondence between
$\varsigma\in\oplus_{a=1}^n\mC[[t_a,t_a^{-1}]\p_{t_a}$ on $\bigcup\limits_{a=1}^nD_a^\times$ and
the Beltrami dif\/ferential $\mu$.
Let
$\varsigma_{\rm out}\in\Gamma_{(\Sigma_g\setminus \vec{z})}$, $\varsigma_{\rm int}\in\oplus_{a=1}^n\mC[[t_a]]\p_{t_a}$.
Consider two equations on $\bigcup\limits_{a=1}^nD_a^\times$
\[
\bar \p\varsigma_{\rm out}=\mu,\qquad \bar \p\varsigma_{\rm int}=\mu,
\]
where $\bar \p|_{D_a^\times}=\p_{\bar t_a}$.
On $D_a^\times \bar\p(\varsigma_{\rm out}-\varsigma_{\rm int})=0$ and, therefore, $\varsigma_{\rm out}-\varsigma_{\rm int}$
represents a Dolbeault cocycle.
The f\/irst equation has solutions that can be continued on $\Sigma_g\setminus \vec{z}$
and the second -- on $\bigcup\limits_{a=1}^nD_a$.
If
$\varsigma\in\oplus_{a=1}^n\mC[[t_a,t_a^{-1}]\p_{t_a}$ has continuations $\varsigma_{\rm out}$ and $\varsigma_{\rm int}$
then it corresponds to a trivial element of $T\gM_{g,n} $.
On the other hand, $\bar \p\varsigma=\mu$ globally and, therefore, $\mu$
represents an exact Dolbeault cocycle.
In this way the non-trivial vector f\/ields $\varsigma\in
\oplus_{a=1}^n\mC[[t_a,t_a^{-1}]\p_{t_a}$ correspond to elements of
$H^1(\Sigma_g, \Gamma(\Sigma_g\setminus  \vec z)\otimes \bar K)$.

\subsection{Def\/inition of conformal blocks and coinvariants}

Let us associate with $\Sigma_{g,n}$ the following set: integer $k$
and the weights $\vec{\mu}=(\mu_1,\ldots,\mu_n$, $\mu_a\in
I_k$) attached to the marked points $\vec{z}=(z_1,\ldots,z_n)$.
The $\hat L(\gg)$-module (\ref{hl})
\begin{gather}\label{reps}
{\bf\hat V}^{[n]}_{\vec{z},\vec{\mu}}=\bigotimes_{a=1}^n{\bf\hat V}_{\mu_a},
\qquad
\vec{z}=(z_1,\ldots,z_n).
\end{gather}
According to (\ref{img})
\begin{gather}\label{hmu}
{\bf \hat V}_{\mu_a}=\bigoplus_{\zeta_a\in\clZ}\hat V_{\mu_a}(\zeta_a).
\end{gather}
Coming back to (\ref{dcc}) we def\/ine a Lie algebra $\gg_{\rm out}=\operatorname{Lie}(G(\Sigma_{g}\setminus  D)$
as a Lie algebra of meromorphic functions on
$\Sigma_{g,n}$ with poles at $\vec{z}=(z_1,\ldots,z_n)$ taking values in $\gg$.
Let $(t_1,\ldots,t_n)$ are local coordinates in
$D$.
There is a homomorphism $\clO(\Sigma_g\setminus  \vec{z})\to\mC[[t_a,t_a^{-1}]$
for each $z_a$ providing the homomorphism of the
Lie algebras $\gg_{\rm out}\to \gg\otimes\mC[[t_a,t_a^{-1}]$.
In this way $\gg_{\rm out}$ acts on $ {\bf\hat
V}^{[n]}_{\vec{z},\vec{\mu}}$ as
\[
(X\otimes f)\cdot(v_1\otimes\cdots\otimes v_n)
=\sum_av_1\otimes\cdots\otimes(X\otimes f(t_a))\cdot v_a\otimes\cdots\otimes v_n.
\]
This is a Lie algebra action.
Due to the residue theorem this homomorphism is lifted to the diagonal central extension
\[
\gg_{\rm out}\hookrightarrow\bigoplus_{a=1}^n\hat L_a(\gg),
\qquad
\hat L_a(\gg)=\left(\gg\otimes\mC[[t_a,t_a^{-1}]\right)\oplus K,\qquad K\to k.
\]

In what follows we need a relation of $\hat V_\mu$ with the space of coinvariants.
In general setting the coinvariants are def\/ined in the following way.
Let $\clW$ be a module of a Lie algebra $\gk$.
The
space of coinvariants $[\clW]_\gk$ is the quotient-space
$[W]_\gk=W/\gk\cdot W$.
In the case at hand we def\/ine
\emph{the space of coinvariants} with respect to the action of $\gg_{\rm out}$,
\[
\clH(\vec{z},\vec{\mu})=\big[{\bf \hat V}^{[n]}_{\vec{z},\vec{\mu}}\big]_{\gg_{\rm out}},
\qquad
\big([V]_\gg=V/\gg\cdot V\big).
\]

\emph{The space of conformal blocks $\clC\big({\bf\hat V}^{[n]}_{\vec{z},\vec{\mu}}\big)$
is the dual space to the coinvariants}.
In other words, $\clC\big({\bf\hat V}^{[n]}_{\vec{z},\vec{\mu}}\big)$ is the space of linear functionals on
$\hat V^{[n]}_{\vec{z},\vec{\mu}}$, invariant under $\gg_{\rm out}$:
\begin{gather*}
F: \ {\bf\hat V}^{[n]}_{\vec{z},\vec{\mu}}\to \mC,
\qquad
F(X\cdot v)=0\quad \text{for~any}\quad X\in\gg_{\rm out}.
\end{gather*}
Put it dif\/ferently, the conformal blocks are $\gg_{\rm out}$-invariant elements of the
contragradient module ${\bf\hat V}_{\vec{z},\vec{\mu}}^{*[n]}$.
For a single marked point case the conformal blocks are
$\gg_{\rm out}$ invariant sections of the line
bundle $\clL_\mu$ over the af\/f\/ine f\/lag variety~(\ref{lbg}).

According to (\ref{reps}) and (\ref{hmu}) the space $ {\bf\hat V}^{[n]}_{\vec{\mu}}$ has the representation
\begin{gather*}
{\bf\hat V}^{[n]}_{\vec{z},\vec{\mu}}= \bigotimes_{a=1}^n\bigoplus_{\zeta_a\in \clZ}
\hat V_{\mu_a}(\zeta_a).
\end{gather*}
In a similar way the conformal blocks are decomposed in subspaces
corresponding to the cha\-racteristic classes of the bundles
\begin{gather*}
\gC\big({\bf\hat V}_{\vec{z},\vec{\mu}}^{[n]}\big)=
\bigotimes_{a=1}^n\bigoplus_{\zeta_a\in \clZ}\gC_a\big(\hat V_{\mu_a}(\zeta_a)\big),
\end{gather*}
where
\begin{gather}\label{scb}
\gC_a=\big\{F(\zeta_a):\hat V_{\mu_a}(\zeta_a)\to\mC\big\}.
\end{gather}

\subsection{Variation of the moduli space of complex structures}

The space of conformal blocks $\gC\big({\bf\hat V}_{\vec{z},\vec{\mu}}^{[n]}\big)$ is a bundle over $\gM_{g,n}$.
This bundle is
equipped with the KZB connection that can be described as follows.

A stress-tensor $T(z,\bar{z})$ in general theories, def\/ined on a surface $\Sigma_{g,n}$,
generates vector f\/ields on $\Sigma_{g,n}$.
A dual object to $T(z,\bar{z})$ is the Beltrami dif\/ferential $\mu(z,\bar{z})$.
It means that there is a~connection on the bundle of
f\/ields over $\clM_{g,n}$
(the Friedan--Shenker connection)
\begin{gather}\label{dol}
\nabla_\mu F=\delta_\mu F+\int_\Sigma \mu T F.                                                    
\end{gather}
In conformal f\/ield theories the stress-tensor is a meromorphic projective structure on $\Sigma_{g,n}$.
The connection acting
on the space of conformal blocks is
projectively f\/lat. The conformal blocks are horizontal
sections of this bundle.
The horizontality conditions are nothing else but the KZB equations for the conformal blocks.
In
general setting these equations are discussed in~\cite{Hi} (for the smooth curves) and in~\cite{Fe}.
They have the form of
non-stationary Schr\"{o}dinger equations~\cite{I}.

The connection (\ref{dol}) can be rewritten in a local form based on the representation (\ref{dccu}).
Let $\bigcup\limits_{a=1}^nD_a^\times\subset\Sigma_g$ and $\gamma_a\subset D_a^\times$
is a small contour and $\varsigma_a$ is a vector f\/ield in $D_a^\times$.
Then
(\ref{dol}) can be written as
\begin{gather}\label{dol1}
\nabla_{\varsigma_a}F=\p_{\varsigma_a}F+\oint_{\gamma_a}\varsigma_a TF
\end{gather}
and the KZB equation assumes the form
\begin{gather}\label{kzb}
\nabla_\varsigma F=0.
\end{gather}
At the marked points $T$ has the second order poles, while
$\varsigma_a\in\mC[[t_a]]\p_{t_a} $ (\ref{dccu}).
Thereby, this integral produces
$\p_{z_a}F$.
On the other hand, the product $TF$ is non-singular outside the disks~$D_a$.
Then for
$\varsigma_a\in\Gamma_{(\Sigma_g\setminus \vec{z})}$ the integrals vanish.
It means that the conformal blocks~$F$ are def\/ined on~$\gM_{g,n}$.

Consider a one point case and let $t$ be a local coordinate on a punctured disk $D^\times$.
The stress-tensor in the local
coordinate has the Fourier expansion
$T(t)=\sum\limits_{n\in\mZ}L_nt^{-n-2}$.
The coef\/f\/icients obey the Virasoro commutation relations
$[L_n,L_m]=(n-m)L_{n+m}+\frac{c}{12}n(n^2-1)$.

In the WZW model the stress-tensor is obtained from the currents by means of the Sugawara construction (see \cite{BF}).
Let $\{\gt_\alpha\}$ be a basis in $\gg$, $\{\gt^\beta\}$ is the dual basis, and
$I_\alpha(t)=\sum_m\gt_{\alpha, m}t^{-m-1}\in\gg\otimes\mC[[t,t^{-1}]$.
Then
\[
T(t)=\frac1{2(k+h^\vee)}\sum_\alpha {:}I_\alpha(t)I^\alpha(t){:},
\]
where $h^\vee$ is the dual Coxeter number.
The Fourier coef\/f\/icients of $T(t)$ take the form
\begin{gather}\label{vc}
L_m=\frac1{2(k+ h^\vee)}\sum_{p\in\mZ}{:}\gt_{\alpha,-p}\gt_{p+m}^\alpha{:}.
\end{gather}
The normal ordering means placing to the right $\gt_{ n}^\alpha$ ($\gt_{\alpha, n}$) with $n>0$.
The Virasoro central charge is
\[
c=\frac{\dim\gg}{k+h^\vee}.
\]
The Virasoro algebra acts on $\gg\otimes\mC[[t,t^{-1}]$ as
\begin{gather}\label{vre}
L_n\mapsto t^{n+1}\frac{d}{dt}.
\end{gather}
This action is well def\/ined because the action of the Sugawara tensor is well def\/ined on the integrable modules.
In particular, it follows from~(\ref{vre}) that for the moving points equation~(\ref{kzb}) assumes the form
\begin{gather}\label{mpo}
\big(\p_{z_a}-L^a_{-1}\big)F=0.
\end{gather}
The restriction of $\nabla_\varsigma$ on $\gC_a$ (\ref{scb})
yields
a family of the KZB equations
\begin{gather}\label{kzbj}
\sum_a\nabla_aF(\zeta_a)=0.
\end{gather}
In next section we construct these equations explicitly for the bundles over elliptic curves.

\subsection{Variation of the moduli space of holomorphic bundles}

\subsubsection{General construction}

The moduli space of holomorphic bundles $\clM_{G,g,n}=\operatorname{Bun}_G$ (\ref{dcc}) is foliated over the moduli space of complex
structures $\gM_{g,n}$.
Let us consider the dependence of the space of coinvariants $\clH(\vec{z},\vec{\mu})$
(conformal blocks $\gC(V^{[n]})$) on
the variations of the moduli of the bundles $\operatorname{Bun}_G$.
For simplicity consider the one-point case.
Let $t_a$ be a local coordinate in $D^\times_a$, and
$G_{\rm out}=G(\Sigma_g\setminus  z_a)$. Def\/ine the quotient
\begin{gather}\label{clm}
M_G=G_{\rm out}\setminus  G(D^\times_a),
\qquad
G(D^\times_a)=G[[t_a,t^{-1}_a].
\end{gather}
This space is the moduli space of $G$-bundles with a trivialization around $z_a$
(see (\ref{dcc})).

Let $\hat V_{\mu_a}$ be an integrable module (\ref{hV}) attached to $z_a$.
Recall that $\hat V^*_{\mu_a}$ is the space of
holomorphic sections $\Gamma(\clL_{\mu_a})$
of the line bundle (\ref{lbg}) over the af\/f\/ine f\/lag variety (\ref{bc}).
In these terms
the space of conformal blocks has the following interpretation \cite{BL,F}.
Since $\hat V_{\mu_a}$ is the integrable
representation, the group $G(D^\times)$ acts on~$\hat V_{\mu_a}$.
Thereby, the subgroup $G_{\rm out}$ acts
on~$\hat V_{\mu_a}$ also.
Due to (\ref{clm}) $G(D^\times_a)$ acts on $\clM_G$ from the right.
Therefore, $G(D^\times_a)$ acts on the sections $\hat V_{\mu_a}\otimes\clO(M_G)$
of the trivial vector bundle $\hat V_{\mu_a}\times\clM_G$
\begin{gather}\label{gact}
g\cdot v(x)=(gv)(xg),
\qquad
g\in G[[t_a,t^{-1}_a],
\qquad
v\in\hat V_{\mu_a}.
\end{gather}
Consider the space of the coinvariants
\[
\hat V_{\mu_a}\otimes\clO(M_G)\big/\big(\hat V_{\mu_a}\otimes\clO(M_G)\big)\cdot \operatorname{Stab}_x,
\]
where $\operatorname{Stab}_x$ is $\operatorname{Lie}(G_x(D^\times)$, $G_x(D^\times)=\{g|x\cdot g=x$, $x\in\clM_G\}$.
In particular,
$\operatorname{Stab}_x=\gg_{\rm out}$ for $x$ corresponding to $G_{\rm out}$.
The spaces of coinvariants are isomorphic for dif\/ferent choices of~$x$.
The dual space $\Gamma(\clL_\mu)/G_{\rm out}$ is the space of conformal blocks.
The quotient $\Gamma(\clL_\mu)/G_{\rm out}$ is a space of sections of the line bundle over $\operatorname{Bun}_G$ (\ref{dcc}).
It means that the space of conformal blocks is a~non-Abelian generalization of the
theta line bundles over the Jacobians.

\subsubsection[$\SLT$-bundles over $\mC P^1$]{$\boldsymbol{\SLT}$-bundles over $\boldsymbol{\mC P^1}$}

It is instructive to consider this construction for $\Sigma=\mC P^1=\mC\cup\infty$.
This case was analyzed in details in~\cite{Bea} for the trivial $G$-bundles and $\gamma=0$ in~(\ref{bd}).
Here we consider $G=\SLT$-bundles with $\gamma\in P^\vee$.
Let $\gt_\alpha=h,e,f$ be the Cartan--Chevalley basis in the
Lie algebra $\sl2$
\[
[h,e]=2e,\qquad [h,f]=2f,\qquad [e,f]=h,
\]
and $\gt_\alpha(n)=\gt_\alpha t^n$.
The Verma module $\clV_\mu$ is generated by $L^-(\sl2)=c\cdot f+\gg\otimes t^{-1}\mC[t^{-1}]$
(\ref{um}), (\ref{lgm}).
The HV $v_\mu$ with the weight $\mu\in P(\sl2)$ is def\/ined by the
conditions $hv_\mu=2s v_\mu$,
$s=\frac12\lan\mu,h\ran\in\frac12\mZ$,
$ev_\mu=0$, $\gt_{\alpha}(n)v_\mu=0$ for $n>0$.
The singular vector $(et^{-1})^{k+1-2s}$ generates the submodule $\clS_\mu\subset\clV_\mu$,
and the integrable module is the quotient $\hat V_\mu=\clV_\mu/\clS_\mu$.

This form of $\hat V_\mu$ def\/ines a trivial sector in (\ref{img}).
Note that $\clZ(\SLT)=\mZ_2=\{\zeta=(0,1)\}$.
Therefore, there are two sectors in the integrable module (\ref{img}).
Consider the non-trivial sector corresponding to $\zeta=1$.
Let $W_P=\{\hat w\}$ be the Weyl group (\ref{awg}), $\hat w=\mZ_2\ltimes t^\gamma$,
where $\gamma$ belongs to the weight lattice $\gamma\in P^\vee(\sl2)=P(\sl2)=\frac12\mZ$.
Since $Q^\vee(\sl2)=Q(\sl2)$,
$\clZ=P/Q\sim\mZ_2$, and $\zeta=1$ corresponds
to $\gamma\notin Q$.
It means that $\lan\gamma,h\ran$ is odd.
Then according to (\ref{img})
\[
\clV_{\mu}(\zeta=1)=\clU\big(L^-(\sl2)\big)(t^\gamma v_\mu),\quad \gamma\notin Q,
\qquad
\hat V_\mu(\zeta=1)=\clV_{\mu}(\zeta=1)/\clS_\mu.
\]
As it was mentioned above, $t^\gamma v_\mu$ is not the HWV.
In other words, it is not annihilated by the positive nilpotent
loop subalgebra (\ref{pns}) $\operatorname{Lie}(N^+(\SLT))=\{n(t)=b\cdot e+\gg \otimes t\mC[[t]]$, $b\in\mC\}$.
In fact, we have
\[
n(t)t^\gamma v_\mu=t^\gamma \operatorname{Ad}^{-1}_{t^\gamma}(n(t)) v_\mu.
\]
Let $\lan\gamma,h\ran=2s>0$, and $s\in\frac12+\mZ$.
Then
\[
\operatorname{Ad}^{-1}_{t^\gamma}(n(t))=\sum_{m\geq 0}\left(a_{m+1}\cdot ht^{m+1}+b_m\cdot et^{2s+m}+c_{m+1}\cdot ft^{-2s+m+1}\right).
\]
The terms $c_{m+1}\cdot ft^{-2s+m+1})$
for $m\leq 2s-1$ do not belong to $\operatorname{Lie}(N^+(\SLT))$.
Multiply $t^\gamma$ by
$wt^{\gamma_1}\in W_Q$, where $\lan\gamma_1,h\ran=-2s+1$ and $w:e\leftrightarrow f$.
This transformation preserves the sector.
Now $n(t)$ annihilates the vector $wt^{\gamma_1}t^\gamma v_\mu$.
Note $(wt^{\gamma_1})$ is uniquely def\/ined by $\gamma$.
Thus, for any
vector $t^\gamma v_\mu$ we def\/ine a unique HWV from the same sector.
It is a particular case of general theorem proved in
\cite{FGK}.

Consider the trivial $G$-bundles over
$\mC P^1$.
It was proved in \cite{Bea} that the conformal blocks are $G$-invariant functionals on
the module $\hat V_\mu^{[n]}$ satisfying some additional conditions.
In particular, for $n=1$
\begin{gather}\label{coin}
\dim(\gC(G))=
\begin{cases}
0, & \mu\neq 0, \\
1, & \mu=0,
\end{cases}
\end{gather}
and for $\hat V_{(0,\infty)(\mu_0,\mu_\infty)}^{[2]}$
\[
\dim(\gC(G))=
\begin{cases}
0, & \mu_0\neq \mu^*_\infty, \\
1, & \mu_0=\mu^*_\infty.
\end{cases}
\]

Let us analyze the case of $\SLT$-bundles.
It follows from the Bruhat decomposition (\ref{PS}) that there are two types of
the $\SLT$-bundles over $\mC P^1$~-- the trivial, when $\gamma\in Q$ in (\ref{grm})
is an element of the root lattice $\gamma\in Q$,
and non-trivial, when $\gamma\notin Q$.
Note that the stable bundles correspond $\gamma=0$.
In the f\/irst case we deal with the adjoint bundles that can be lifted to the
$\SLT$-bundles. In the second case there is an obstruction to lift these bundles to the
$\SLT$-bundles.

Let $z^{-1}$ be a local coordinate in a neighborhood of $\infty$.
The Lie algebra $\gg_{\rm out}$ assumes the form
$\gg_{\rm out}=\sl2+z^{-1}\sl2\otimes\mC[z^{-1}]$.
Let $n=1$ and $z=0$ is the marked point with the attached integrable
$L(\sl2)$-module $\hat V_\mu$.
We have $\gg_{\rm out}(v_\mu)=\hat V_\mu$ for $\mu\neq 0$.
But $v_0\notin\gg_{\rm out}(v_0)$ and $v_0$
is the coinvariant conf\/irming~(\ref{coin}).

Consider the integrable module generated by $z^\gamma v_\mu$, where, as above, $\lan\gamma,h\ran=2s>0$,
and $s\in\frac12+\mZ$.
Then
\[
\gg_{\rm out}z^\gamma v_\mu= \operatorname{Ad}^{-1}_{z^\gamma}(n(t))
=z^\gamma \sum_{m\geq 0}\big(a_{-m}z^{-m}\cdot h+b_{-m}z^{-2s-m}\cdot e+c_{-m}z^{2s-m}\big)\cdot fv_\mu.
\]
Then the elements $b_{-m}z^{-2s-m}\cdot e(v_\mu)$ for $0\leq m<2s$ are not generated by $\gg_{\rm out}$.
Thus, if $\gamma\neq 0$,
the space of coinvariants (and the space of conformal blocks) is non-empty
for an arbitrary weights $\mu$.
Its dimension depends on $\gamma$:
$\dim(\gC(\SLT))=2s$ (compare with (\ref{coin})).

\subsubsection{The form of connection}

For conformal blocks we have (see (\ref{gact}))
\begin{gather*}
F(x)=(gF)(xg),\qquad g\in \operatorname{Stab}_x.
\end{gather*}

Def\/ine the current
$J(t_a)=(g^{-1}dg)(t_a)\in\gg\otimes\mC[[t_a,t_a^{-1}])\otimes\Omega^1(D_a^\times)$
for $g(t_a)\in G\otimes\mC[[t_a,t_a^{-1}]$.
A local version of (\ref{gact}) is def\/ined by the operator
\begin{gather*}
\nabla_{\bfu_\alpha}=\p_{\bfu_\alpha} +\oint_{\gamma_a}\lan\big(g^{-1}dg\big),\gt_\alpha\ran,
\end{gather*}
where $\gt_\alpha$ is a generator of $\gg$ and $\bfu_\alpha$ is a coordinate of the tangent vector to $\operatorname{Bun}_G$.
The action of $\nabla_{\bfu_\alpha}$ on the conformal blocks is well def\/ined because the conformal blocks are
$\gg_{\rm out}$-invariant.
Therefore, they are horizontal with respect to this connection
\begin{gather}\label{cms1}
\p_{\bfu_\alpha}F +\oint_{\gamma_a}\lan\big(g^{-1}dg\big)(t_a),\gt_\alpha\ran F=0.
\end{gather}
If one takes $\bfu$ from $\clM^{(\zeta)}_{G,g,n}$ (\ref{grm1}) then (\ref{cms1}) takes the form
\begin{gather*}
\p_{\bfu_\alpha(\zeta_a)}F(\zeta_a) +\oint_{\gamma_a}\lan\big(g^{-1}dg\big)(t_a),\gt_\alpha\ran F(\zeta_a)=0,
\end{gather*}
where $F(\zeta_a)\in\gC_a$ (\ref{scb}).

\section{KZB equations related to elliptic curves\\ and non-trivial bundles}\label{section4}

\subsection{Moduli space of elliptic curves}

We consider in details the genus one case $\Sigma_{1,n}$.
Let $\Sigma_\tau=\mC/\lan\tau,1\ran$ be the elliptic curve with the
modular parameter in the upper half-plane $\clH=\{\operatorname{Im} m\tau>0\}$.
For $n\in\mZ$, $n\geq 1$ def\/ine the set of marked points
$\vec{z}=(z_1,\ldots,z_n)$.
Due to the $\mC$ action on $\Sigma_\tau (z\to z+c)$, we assume that $\sum_az_a=0$.
A big cell
$\gM^0_{1,n}$ in the Teichm\"uller space $\gM_{1,n}$ is def\/ined as
\begin{gather*}
\gM^0_{1,n}=\left\{(z_1,\ldots,z_n),\; \sum_az_a=0,\; z_k\neq z_j,\, \text{mod}(\lan\tau,1\ran)\right\}
\times\clH.
\end{gather*}

\subsection[Moduli space of holomorphic $G$-bundles over elliptic curves]{Moduli space of holomorphic $\boldsymbol{G}$-bundles over elliptic curves}

For $G=GL_N$ the moduli space of holomorphic bundles was described by M.~Atiyah~\cite{At}.
For the trivial $G$-bundles, where~$G$ is a complex simple group, it was done in \cite{BS1,BS2,Loo}.
Non-trivial $G$-bundles and their moduli spaces were considered in~\cite{FM1,FM2,FMW,Sch}.
We describe the moduli space of stable non-trivial holomorphic bundles over $\Sigma_\tau$
using an approach of~\cite{LOSZ}.

Let $G$ be a complex simple Lie group.
An universal cover $\bar G$ of $G$
in all cases apart $G_2$, $F_4$ and $E_8$ has a non-trivial center $\clZ(\bar G)$.
The adjoint group is the quotient $G^{\rm ad}=\bar G/\clZ(\bar G)$.
For the cases $A_{n-1}$ (when $n=pl$ is non-prime) and $D_n$ the center $\clZ(\bar G)$ has non-trivial subgroups
$\clZ_l\sim\mu_l=\mZ/l\mZ$.
Assume that $(p,l)$ are co-prime.
There exists the quotient-groups
\begin{gather}\label{fgl}
G_l=\bar{G}/\clZ_l,
\qquad
G_p=G_l/\clZ_p,
\qquad
G^{\rm ad}=G_l/\clZ(G_l),
\end{gather}
where $\clZ(G_l)$ is the center of $G_l$ and $\clZ(G_l)\sim\mu_p=\clZ(\bar G)/\clZ_l$.

Following \cite{NS} we def\/ine a
$G$-bundle $E_G=\clP\times_GV$ by the
transition operators $\clQ$ and $\Lambda_j$ acting on the sections of $s\in\Gamma(E_{G})$ as
\begin{gather}\label{solu}
\bfs(z+1)=\clQ(z)\bfs(z),
\qquad
\bfs(z+\tau)=\Lambda(z) \bfs(z),
\end{gather}
where $\clQ(z)$ and $\Lambda(z)$ take values in End$(V)$.
Going around the basic cycles of $\Sigma_\tau$ we come to the
equation
\begin{gather}\label{1eq}
\clQ(z+\tau)\Lambda(z)\clQ(z)^{-1}\Lambda^{-1}(z+1)=\operatorname{Id}.
\end{gather}
It follows from \cite{NS} that it is possible to choose the constant
transition operators.
Then we come to the equation
\begin{gather}\label{1}
\clQ\Lambda\clQ^{-1}\Lambda^{-1}=\operatorname{Id}.
\end{gather}

Replace (\ref{1}) by the equation
\[
\clQ\Lambda\clQ^{-1}\Lambda^{-1}= \zeta \operatorname{Id},
\]
where $\zeta $ is a generator of the center $\clZ\big(\bar{G}\big)$.
In this case $(\clQ,\Lambda)$ are the clutching operators
for $G^{\rm ad}$-bundles, but not for $\bar{G}$-bundles,
and $\zeta$ plays the role of obstruction to lift the $G^{\rm ad}$-bundle
to the $\bar{G}$-bundle.
Here $\zeta=\bfe(\varpi^\vee)$ is a generator of the center $\clZ\big(\bar{G}\big)$, where $\varpi^\vee\in P^\vee$
is a~fundamental coweight such that $N\varpi^\vee\in Q^\vee$
and $N=\operatorname{ord}(\clZ\big(\bar{G}\big))$.\footnote{For the simplicity
we assume here and in what follows that $\clZ\sim\mZ_l$.
The case $\clZ(\operatorname{Spin}(4n))=\mZ_2\oplus\mZ_2$ can be considered in a similar way.}

Let $0< j\leq N$.
Consider a bundle with the space of sections with the quasi-periodicities
\begin{gather}\label{mon1}
s(z+1)=\clQ s(z),
\qquad
s(z+\tau)=\Lambda_j s(z)
\end{gather}
such that
\begin{gather}\label{o01}
\clQ\Lambda_j\clQ^{-1}\Lambda_j^{-1}= \zeta^j \operatorname{Id}.
\end{gather}

If $j$ and $N$ are co-prime numbers then $\zeta^j$ generates $\clZ\big(\bar{G}\big)$.
In this case $\clQ$ and $\Lambda_j$ can serve as transition operators only for a $G^{\rm ad}=\bar{G}/\clZ$-bundle,
but not for
$\bar{G}$-bundle and $\zeta^j$ is an obstruction to lift $G^{\rm ad}$-bundle to $\bar{G}$-bundle.

The element $\zeta$ has a cohomological interpretation.
It is called \emph{the characteristic class} of $E_G$.
It can be
identif\/ied with elements
of the group $H^2(\Sigma_{g,n},\clZ\big(\bar{G}\big))$.
This group classif\/ies the of the characteristic classes
of the bundles \cite{LOSZ}.

Consider, as above, a non-prime $N=pl$ and put
$j=p$.
Then $\zeta^j$ is a
generator of the group~$\mZ_l$.
In this case $\clQ$ and $\Lambda_j$ are transition operators for $G_l=\bar{G}/\clZ_l$-bundles (see
(\ref{fgl})) and~$\zeta^j$ is an obstruction to lift a $G_l$-bundle to a $\bar{G}$-bundle.

The moduli space of stable holomorphic over $\Sigma_\tau$ with the sections~(\ref{mon1}) is def\/ined as
\begin{gather}\label{mos}
\clM^{(j)}_{G,1}=(\text{solutions~of \eqref{o01}})/(\text{conjugation}).
\end{gather}
For the stable bundles this description of the moduli space is equivalent to (\ref{dcc}).
In fact, the monodromy of $\bfs(z)$ around $z=0$
is the same as in (\ref{mon}).
Similar to (\ref{grm1}) we have
\begin{gather}\label{mbe}
\clM_{G,1}=\bigcup_{j=1}^N\clM^{(j)}_{G,1}.
\end{gather}

Assume that $\clQ$ is a semi-simple element and $\clQ\in\clH_{\bar{G}}$ is
a f\/ixed Cartan subgroup of $\bar{G}$.
It means that we consider an open subset
\[
\clM^{(j)}_{G,1}\supset\Big(\clM^{(j)}_{G,1}\Big)^0\equiv\clM^{(j)}_0(G)
=\big\{(\clQ\in\clH_{\bar{G}},\Lambda_j)/(\text{conjugation})\big\}.
\]
In this case the solutions of (\ref{1}) have the form \cite{LOSZ}
\begin{gather}\label{trm}
\clQ=\exp\left(2\pi i\frac{\rho^\vee}{h}\right),
\qquad
\Lambda_j=\Lambda_0V_j,
\end{gather}
where $\rho^\vee$ is a half-sum of positive coroots, $h$ is the Coxeter number,
$\Lambda_0$ is an element of the Weyl group def\/ined by $\zeta^j$:
\[
\zeta^j\to \Lambda_0,
\qquad
(\Lambda_0)^l=\operatorname{Id}.
\]
The element $\Lambda_0$ preserves the extended system of simple roots $\Pi^{\rm ext}=\Pi\cup(\alpha_0)$,
where $-\alpha_0$ is a maximal
root~\cite[Proposition~3.1]{LOSZ}.
In this way
$\Lambda_0$ is a symmetry of the extended
Dynkin diagram of $\gg=\operatorname{Lie}\big(\bar{G}\big)$, generated by $\varpi^\vee$~\cite{Bou}.

Let $\tilde\clH_0\subset\clH_{\bar{G}}$ be the
Cartan subgroup commuting with $\Lambda_0$.
To describe $V_j$ consider the adjoint action
$\lambda=\operatorname{Ad}(\Lambda_0)$ on the Cartan subalgebra $\gh=\operatorname{Lie}(\clH_{\bar{G}})$.
Let $\tilde\gh_0=\operatorname{Lie}(\tilde\clH_0)$ be the invariant subalgebra
($\lambda(\tilde\gh_0)=\tilde\gh_0$).
Then
$V_j=\exp(2\pi\imath\bfu) (\bfu\in\tilde\gh_0)$ is an arbitrary element from~$\tilde\clH_0$
def\/ining the moduli space $\clM^{(j)}_0(G)$.

There exists a basis $\tilde\Pi^{\vee}_j$ in
$\tilde\gh_0$ such that $\tilde\Pi$ is a system of simple roots for a simple Lie subalgebra $\tilde\gg_0\subset\gg$.
For the list of
these subalgebras see \cite{LOSZ}.
If $j=N$, we come to the trivial bundles (\ref{1}).
In this case $\Lambda_0=\operatorname{Id}$, $\tilde\gh_0=\gh$ and $\tilde\gg_0=\gg$.

Let $\tilde{Q}^\vee$ and $\tilde{P}^\vee$ be the coroot and the coweight lattices
in $\tilde\gh_0$, and $\tilde{W}$ is the Weyl group corresponding to $\tilde\Pi$.
Def\/ine the Bernstein--Schwarzman type groups
\cite{BS1,BS2}.
They are constructed by means of the lattices $\tilde{Q}^\vee$ or $\tilde{P}^\vee$.
In the f\/irst case it is the semidirect products
\begin{gather}\label{twsc}
\tilde W_{\rm BS}=\tilde W\ltimes\big(\tau \tilde Q^\vee\oplus \tilde Q^\vee\big).
\end{gather}
Then the moduli space of non-trivial $\bar G$-bundles with the characteristic class $\zeta^j$
is the fundamental domain in
$\tilde\gh^{(j)}_0$ under the action of $\tilde W_{\rm BS}$
\begin{gather}\label{tsc}
\clM^{(j)}_0(\bar G)=C^{\rm sc}_j=\tilde\gh^{(j)}_0/\tilde W_{\rm BS}
\end{gather}
is~the moduli space of non-trivial $\bar{G}$-bundles.

Consider $G^{\rm ad}$-bundles.
Def\/ine the semidirect product
\begin{gather}\label{twad}
W^{\rm ad}_{\rm BS}=\tilde W\ltimes\big(\tau\tilde P^\vee\oplus \tilde P^\vee\big).
\end{gather}
A fundamental domain of this group in $\tilde\gh^{(j)}_0$ is $C^{\rm ad}=\tilde\gh^{(j)}_0/\tilde W^{\rm ad}_{\rm BS}$
and
\begin{gather}\label{tad}
\clM^{(j)}_0\big(G^{\rm ad}\big)= C_j^{\rm ad}=\tilde\gh^{(j)}_0/\tilde W^{\rm ad}_{\rm BS}
\end{gather}
is the moduli space of the non-trivial $G^{\rm ad}$-bundles.
It is the moduli space of $E_{G^{\rm ad}}$-bundles with characteristic class def\/ined by $\zeta^j$.
In other words
\[
\bfu\in
\begin{cases}
C^{\rm sc}_j &  \text{for~$E_{\bar{G}}$-bundles}, \\
C^{\rm ad}_j &  \text{for~$E_{G^{\rm ad}}$-bundles}.
\end{cases}
\]

\subsection{The gauge Lie algebra for elliptic curves }

Here we def\/ine the moduli space of holomorphic $G$-bundles coming back to the double coset construction (\ref{dcc}).
Recall, that the Lie algebra $\gg_{\rm out}=\operatorname{Lie}(G(\Sigma_{\tau,n}\!\setminus \!\vec{z}))$
is a Lie algebra of meromorphic functions on
$\Sigma_{\tau,n}$ with poles at $\vec{z}=(z_1,\ldots,z_n)$ and the quasi-periodicities~(\ref{solu}),~(\ref{1eq}).

Let us take for simplicity the case (\ref{1}) and apply
the decomposition (\ref{gra}) corresponding the characteristic class def\/ined by $\zeta$ to
the Lie algebra $\gg_{\rm out}$:
\[
\gg_{\rm out}=\oplus_{k=0}^{l-1} \gg_k,
\qquad
\gg_0= \gg'_0\oplus\tilde \gg_0,
\qquad
\operatorname{Ad}_{\Lambda_0}( \gg_k(z))=\bfe\left(\frac{k}l\right) \gg_k(z).
\]
Consider the quasi-periodicity conditions (\ref{solu}).
The GS-basis is diagonal under $\operatorname{Ad}_\Lambda$ and $\operatorname{Ad}_\clQ$ actions
(\ref{qc1})--(\ref{qr}).
We should f\/ind functions on $\Sigma_\tau\setminus  D$ that have
the same phase-factors and pole
singularities at $\vec{z}$.
To def\/ine $\gg_k$ and $\gg_0'$ we use the functions $\phi$ (\ref{phi}),
and $\varphi_\alpha^{k,m}$~(\ref{mmfi}).
They have the needed quasi-periodicities (\ref{A.14}), (\ref{A.14a}) and poles at
$z=0$ (\ref{mmfi0}), (\ref{mmfi1}).
Then we f\/ind
\begin{gather}
\gg_k=\Bigg\{\sum_{a=1}^n\sum_{m=0}^{K(a,k)}\Bigg(\sum_{\alpha\in\Pi}
x^k_{\alpha,m,a}\p_z^m\phi\left( \frac{k}l,z - z_a \right) \gh^k_\alpha
+ \sum_{\alpha\in R}y^k_{\alpha,m,a}\varphi_\alpha^{k,m}(\bfu,z-z_a)\gt_\alpha^k
\Bigg)\Bigg\},\!\!\!\label{gouj}\\
\gg'_0=\Bigg\{\sum_{a=1}^n\sum_{\alpha\in R}\sum_{m=0}^{K(a,\alpha)}y'_{\alpha,m,a}
\varphi_\alpha^{0,m}(\bfu,z-z_a)\gt_\alpha^0\Bigg\}.\label{goup}
\end{gather}
Similarly, from (\ref{A.1}), (\ref{A.2}), (\ref{A.29}), (\ref{A.12}), (\ref{A.13}) we have
\begin{gather}
\tilde\gg_0=\Bigg\{\sum_{a=1}^n\Bigg(
\sum_{\alpha\in\tilde\Pi}\Bigg(x^0_{\alpha,0}+ \sum_{m=1}^{K(a,\alpha)} x^0_{\alpha,m,a}E_m(z-z_a) \Bigg)h_\alpha\nonumber\\
 \phantom{\tilde\gg_0=}
+\sum_{\alpha\in\tilde R}\sum_{m=0}^{K(a,\alpha)} y^0_{\alpha,m,a}\varphi_\alpha^{0,m}(\bfu,z-z_a) E_{\alpha}
\Bigg)\Bigg\}.\label{gou0}
\end{gather}
Then $\gg_{\rm out}$ has the correct quasi-periodicities and has poles of orders
$K(a,m)$, $K(a,\alpha)$ at $z_a$, $a=1,\ldots,n$.
In this last expression (due to the residue theorem) from (\ref{A.12}) we assume that
\begin{gather}\label{sop}
\sum_{a=1}^nx^0_{\alpha,1,a}=0.
\end{gather}
Let us unify the last two expression (\ref{goup}) and (\ref{gou0}) in a single formula,
\begin{gather}
\gg_0=\Bigg\{\sum_{a=1}^n\Bigg(
\sum_{\alpha\in\tilde\Pi}\Bigg(x^0_{\alpha,0}+ \sum_{m=1}^{K(a,\alpha)} x^0_{\alpha,m,a}E_m(z-z_a) \Bigg)
h_\alpha\nonumber\\
 \phantom{\gg_0=}
+\sum_{\alpha\in R}\sum_{m=0}^{K(a,\alpha)} y^0_{\alpha,m,a}\varphi_\alpha^{0,m}(\bfu,z-z_a)\gt^0_{\alpha}
\Bigg)\Bigg\}.\label{gou}
\end{gather}

We will act on the coinvariants by $\gg_{\rm out}$.
In what follows we need the limit
$z\to z_a$ of these expressions.
Notice that $\gg_{\rm out}$ is the f\/iltered Lie algebra.
The f\/iltration is def\/ined by the orders of poles.
The behavior of $\gg_{\rm out}$
is def\/ined by the asymptotics
(\ref{mmfi0})--(\ref{mmfi1}), (\ref{A.29}).
As it will become clear below we need the least singular terms in $\gg_{\rm out}$.
In
this way we take $m=0$ in (\ref{gouj}), (\ref{gou}) and $m=1$ $(E_1(z-z_a))$ in (\ref{gou}):
\begin{gather}
\gg_k\sim\sum_{a=1}^n\sum_{\alpha\in\Pi}x^k_{\alpha,0,a}
\Bigg(\cdots+\frac1{z-z_a}+E_1\left(\frac{k}l\right)
+\sum_{b\neq a}\phi\left(\frac{k}l,z_a-z_b\right)+\Bigg)\gH^k_\alpha\nonumber\\
 \phantom{\gg_k\sim}
+y^k_{\alpha,0,a}\Bigg(
\cdots+\frac1{z-z_a}+E_1\left(\lan\bfu+\kappa\tau,\alpha\ran+\frac{k}l\right)+2\pi\imath\lan\kappa,\alpha\ran\nonumber\\
 \phantom{\gg_k\sim}
+\sum_{b\neq a}\varphi_\alpha^{k,0}(\bfu,z_b-z_a) +\cdots
\Bigg)\gt_\alpha^k,\label{as}
\\
\gg_0\sim \sum_{\alpha\in\tilde\Pi}
\Bigg(\cdots+x^0_{\alpha,1,a}\Bigg(\frac1{z-z_a}+\sum_{b\neq a}E_1(z_b-z_a)\!\Bigg)+x^0_{\alpha,0}+\cdots\Bigg)h_\alpha\nonumber\\
 \phantom{\gg_0\sim}
+\sum_{\alpha\in\tilde R}y^0_{\alpha,0,a}\Bigg(\cdots+\frac1{z-z_a}
+E_1\big(\lan \bfu+\kappa\tau,\alpha\ran\big)+2\pi\imath\lan\kappa,\alpha\ran\nonumber\\
 \phantom{\gg_0\sim}
+\sum_{b\neq a} \varphi^0_\alpha(\bfu,z_b-z_a)+\cdots\Bigg)\gt^0_{\alpha}.\label{as1}
\end{gather}
Here ``$\cdots$'' means the terms of order $o\big(z-z_a\big)^{-1}$ and $o(1)$.
For $\gg_{\rm int}=\operatorname{Lie}(G(U_D))$
we have local expansions
in neighborhoods of the marked points
\[
\gg_{\rm int}=\Bigg\{X=\sum_{a=1}^n\Bigg(b_a +\sum_{j>0} y^a_j(z-z_a)^j\Bigg),
\  b_a\in\gb_0,\  y^a_j\in\gg\Bigg\}.
\]
Def\/ine the Lie algebra with the loose condition (\ref{sop}))
\[
\gg'_{\rm out}=\gg_{\rm out}\qquad\text{with}\quad\sum_{a=1}^nx^0_{\alpha,1,a}\in\mC
\]
and let $\gn^-=\sum\limits_{\alpha\in R^+}\gg^{-\alpha}$.
Then the Lie algebra $\operatorname{Lie}(G(D^\times))$ has the form (compare with the general
case (\ref{lms}))
\begin{gather}
\operatorname{Lie}\big(G(D^\times)\big)=\gg'_{\rm out}\oplus\big(\oplus_{a=1}^n\gn^{-}_a\big)\oplus\gg_{\rm int}\nonumber\\
\hphantom{\operatorname{Lie}\big(G(D^\times)\big)}{}
=\gg_{\rm out}\oplus\Bigg(\sum_{a=1}^n\sum_{\alpha\in\tilde\Pi}x^0_{\alpha,1,a}h_\alpha\Bigg)
\oplus\big(\oplus_{a=1}^n\gn^{-}_a\big)\oplus\gg_{\rm int}.\label{ldcc}
\end{gather}
Notice that the constant terms $\gn^{-}_a$ come from the constant terms $c(m,k)$ in (\ref{mmfi1}).
We can conclude from
(\ref{ldcc}) that locally the action on
$G(D^\times)$ by $G_{\rm out}=
G(\Sigma_{1,n}\setminus \vec{z})$ from the left and by $G_{\rm int}=\prod\limits_{a=1}^nG(D_a)$
from the right absorbs almost all negative
and positive modes of $G(D^\times)$ except the two types of modes describing the moduli space:
\begin{itemize}\itemsep=0pt
\item The vector $\bfu=\sum\limits_{a=1}^n\sum\limits_{\alpha\in\tilde\Pi}x^0_{\alpha,1,a}h_\alpha\in\tilde\gh_0$.
It def\/ines an element of
the moduli space~$\clM_{G,1}$~(\ref{mbe}).
\item The Lie algebras $\gn^{-}_a$, $a=1,\ldots n$.
They are the tangent spaces to the f\/lag varieties attached
the marked points coming from the quasi-parabolic structure of the bundle.
\end{itemize}

\subsection{Conformal blocks}

In this section we def\/ine connections on
the space of conformal blocks and derive the KZB equations in a similar
way as it was done for the trivial characteristic classes in \cite{FW}.
The derivation is based on the representation of the
moduli space of bundles as the double coset space (\ref{dcc}) in a
given sector of the decomposition (\ref{grm}).
In other words, the characteristic class (def\/ined by $j=0,\ldots, l-1$ in
(\ref{mos})) is f\/ixed and we deal with
$G_{\text{\rm out}} \setminus  G^{\rm ad}_\gamma\otimes\mC[[t_a,t_a^{-1}]/G_{\rm int}$, where
$G_{\text{\rm out}}=G(\Sigma_{\tau,n}\setminus  \vec{z})$ and $G_{\rm int}=G\otimes\mC[[t_a]]$ were described above.

Let us write down the Virasoro generators (\ref{vc}) using the GS-basis (\ref{ft}), (\ref{cb}), (\ref{dbh})
$\gt_\alpha^k\otimes t_a^{m} \equiv \gt_\alpha^k(m)$ ($\gt_\alpha^k(0) \equiv \gt_\alpha^k$)
for the generators of the loop algebra
\begin{gather}\label{Vir1}
L_{m}^a=\frac1{2(k + h^\vee)}\sum_{p\in\mZ}\sum_{q=0}^{l-1}
\Bigg(\!\sum_{\alpha\in R}{:}|\alpha|^2\gt_\alpha^q(-p)\gt_{-\alpha}^{-q}(p+m){:}+
 \sum_{\alpha\in \tilde\Pi} {:}\gH_\alpha^q(-p)\gh_{\alpha}^{-q}(p+m){:}
 \!\Bigg).\!\!\!
\end{gather}
Consider the integrable modules attached to the marked points $\hat V^{[n]}_{\vec{z}\vec{\mu}}$
(\ref{reps}) and the
corresponding conformal blocks.
They satisfy the equations (\ref{kzb}), (\ref{kzbj}), (\ref{cms1}).
For elliptic curve they
assume the form:
\begin{itemize}\itemsep=0pt
\item The moving points (\ref{mpo}):
\begin{gather}\label{kze1}
\big(\p_a-L_{-1}^a\big)F=0,
\qquad
 \p_a=\frac1{k+h^\vee}\p_{z_a},  \qquad   t_a=z-z_a .
\end{gather}
\item The vector f\/ield corresponding to the deformation of the moduli $\tau$ of
the elliptic cur\-ve~$\Sigma_{\tau,n}$:
\begin{gather}\label{kze2}
\left(\p_\tau-\frac{1}{2\pi i}E_1(z)\p_z\right)F=0.
\end{gather}
This action follows from (\ref{dol1}) and the operator algebra
\[
T(z')F(z)=E_1(z'-z)\p_zF(z)+ \text{analitic~part}.
\]
\item The invariance with respect to the action of $\gg_{\rm out}$ (\ref{cms1}):
\begin{gather}\label{kze12}
\left(l\p_{\bfu_{-\alpha}}+E_1(z)\gH_\alpha^{0}\right)F=0,\quad \alpha\in\tilde\Pi,\qquad \bfu=\{\bfu_{\alpha}\},
\quad
\alpha\in\tilde\Pi,
\end{gather}
where $\gH_\alpha^0$ are the Cartan generators~(\ref{dbh}).
Notice that this operator is well def\/ined
on~$M_G$~(\ref{clm}).
\end{itemize}

The vector f\/ield (\ref{kze2}) is def\/ined on the universal curve
$\clH\times\mC/\lan\tau,1\ran\setminus \clH\times 0$,
since it is invariant under the lattice shifts $\lan\tau,1\ran$.
The $\tau$ deformation can be def\/ined in the non-holomorphic form as $\p_\tau+\frac{z-\bar{z}}{\tau-\bar\tau}\p_z$.

The invariance with respect to $\operatorname{Lie}(G_{\text{\rm out}})$ (\ref{gouj}), (\ref{gou}) means that
\begin{gather}
\varphi_\alpha^k(\bfu_\alpha,z-z_a)\gt_\alpha^k F=0,\qquad\alpha\in R,\quad \forall\, k,\nonumber\\
\varphi_\alpha^k(0,z-z_a)\gH_\alpha^k F=0,\qquad \alpha\in\tilde\Pi,\quad k\neq 0.\label{kze11}
\end{gather}
Now using (\ref{as}), (\ref{as1}) and (\ref{kze11}) we write down the annihilation condition $\gg_{\rm out}F=0$ in
in the basis
$\gt_\alpha^{k,c}(m)=1\otimes\cdots\otimes 1\otimes \gt_\alpha^{k,c}(m)\otimes 1\otimes \cdots\otimes 1$ (on the $c$-th place):
\begin{gather}
 \Bigg(\gt_\alpha^{k,a}( - 1)+
\left(E_1\left(\bfu_\alpha+\lan\kappa,\alpha\ran\tau+\frac{k}l\right)+2\pi i\lan\kappa,\alpha\ran\right)
\gt_\alpha^{k,a}(0)\nonumber\\
 \qquad{} +\sum\limits_{c\neq a}\varphi_\alpha^{k}(\bfu_\alpha,z_c-z_a)\gt_\alpha^{k,c}(0) \Bigg)F=0,\nonumber
\\
 \Bigg(\gH_\alpha^{k,a}( - 1)
+\left(E_1\left(\lan\kappa,\alpha\ran\tau+\frac{k}l\right)+2\pi i\lan\kappa,\alpha\ran\right)
\gH_\alpha^{k,a}(0)\nonumber\\
 \qquad{} +\sum\limits_{c\neq a}\varphi_\alpha^{k}(0,z_c-z_a) \gH_\alpha^{k,c}(0) \Bigg)F=0\label{kze111}
\end{gather}
for $\alpha\in R$, $\forall\, k$ and $\alpha\in \tilde\Pi$, $k\neq 0$ correspondingly.
In the same way (\ref{kze2}) and (\ref{kze12})
assume the form
\begin{gather}
\Bigg(\p_\tau+\frac{1}{2\pi i}L_{-2}^a+\frac{1}{2\pi i}\sum\limits_{c\neq a} E_1(z_c-z_a)L_{-1}^a\Bigg)F=0,
\nonumber\\
\Bigg(l\p_{\bfu_{-\alpha}}+\gH_\alpha^{0,a}( - 1)+\sum\limits_{c\neq a}E_1(z_c-z_a)\gH_\alpha^{0,c}(0) \Bigg)F=0,
\qquad
\alpha\in\tilde\Pi.\label{kze121}
\end{gather}

Now we are ready to evaluate the Virasoro generators, i.e.\
to express them in terms of zero modes of the loop algebra
$\gt_\alpha^{k,c}(0) \equiv \gt_\alpha^{k,c}$ only.
As we have found above the positive modes of the loop algebra act on $F$ by
zero $\gt_\alpha^{k,a}(m)F=0$, $m\in{\mathbb Z}_+$.
Therefore, from (\ref{Vir1}) we have
\begin{gather}\label{Vir2}
(k+h^\vee)L_{-1}^a=\sum\limits_{q=0}^{l-1}
\Bigg(\sum\limits_{\alpha\in R}\gt_\alpha^{q,a}( - 1)\gt_{-\alpha}^{-q,a}(0)
+\sum\limits_{\alpha\in \tilde\Pi}\gH_\alpha^{q,a}( - 1)\gh_{\alpha}^{-q,a}(0)\!\Bigg)\qquad \text{on}~F
\end{gather}
and
\begin{gather}
(k+ h^\vee)L_{-2}^a= \sum\limits_{q=0}^{l-1}
\Bigg(\sum\limits_{\alpha\in R}\gt_\alpha^{q,a}( - 2)\gt_{-\alpha}^{-q,a}(0)+
 \sum\limits_{\alpha\in \tilde\Pi}\gH_\alpha^{q,a}( - 2)\gh_{\alpha}^{-q,a}(0) \Bigg)\label{Vir3}\\
\hphantom{(k+ h^\vee)L_{-2}^a=}{} +\frac12 \sum\limits_{q=0}^{l-1}
\Bigg(\sum\limits_{\alpha\in R}\gt_\alpha^{q,a}( - 1)\gt_{-\alpha}^{-q,a}( - 1)+
 \sum\limits_{\alpha\in \tilde\Pi}\gH_\alpha^{q,a}( - 1)\gh_{\alpha}^{-q,a}( - 1) \Bigg)\qquad \text{on}~F.\nonumber
\end{gather}

In order to f\/ind $L^a_{-1}$ one need to substitute $\gt_\alpha^{k,a}( - 1)$, $\gH_\alpha^{k,a}( - 1)$
from (\ref{kze111})
and $\gH_\alpha^{0,a}( - 1)$ from~(\ref{kze121}) into~(\ref{Vir2})
\begin{gather*}
-(k+h^\vee)L_{-1}^a=l \sum\limits_{\alpha\in \tilde\Pi}\gh_{\alpha}^{0,a}(0)\p_{\bfu_{\alpha}}\\
\hphantom{-(k+h^\vee)L_{-1}^a=}{}
                    + \sum\limits_{q=0}^{l-1}\,\sum\limits_{\alpha\in R}|\alpha|^2
                      \Bigg(\left(E_1\left(\bfu_\alpha+\lan\kappa,\alpha\ran\tau+\frac{q}l\right)
                      +2\pi i\lan\kappa,\alpha\ran\right)\gt_\alpha^{q,a}(0)\gt_{-\alpha}^{-q,a}(0)\\
\hphantom{-(k+h^\vee)L_{-1}^a=}{}
                    + \sum\limits_{c\neq a}\varphi_\alpha^{q}(\bfu_\alpha,z_c-z_a)
                      \gt_\alpha^{q,c}(0)\gt_{-\alpha}^{-q,a}(0)\Bigg)\\
\hphantom{-(k+h^\vee)L_{-1}^a=}{}
                    + \sum\limits_{q=0}^{l-1}\,\sum\limits_{\alpha\in \tilde\Pi}
                      \Bigg(\left(E_1\left(\lan\kappa,\alpha\ran\tau+\frac{q}l\right)+2\pi i\lan\kappa,\alpha\ran\right)
                      \gH_\alpha^{q,a}(0)\gh_{-\alpha}^{-q,a}(0)\\
\hphantom{-(k+h^\vee)L_{-1}^a=}{}
                    + \sum\limits_{c\neq a}\varphi_\alpha^{q}(0,z_c-z_a)\gH_\alpha^{q,c}(0)\gh_{-\alpha}^{-q,a}(0)\Bigg),
\end{gather*}
where $\varphi_\alpha^{0}(0,z_c - z_a)=E_1(z_c - z_a)$.
The f\/irst term in the last line vanishes due to skew-symmetry with
respect to $\alpha,q\rightarrow -\alpha,-q$.
The similar term in the second line does not vanish because
$[\gt_\alpha^{q,a}(0),\gt_{-\alpha}^{-q,a}(0)]
=\frac{p_\alpha}{\sqrt{l}}\exp\left(-2\pi i\frac{q}{l}\right)\gh^{0,a}_\alpha$
\cite{LOSZ}.
Therefore,
\begin{gather*}
 \sum\limits_{q=0}^{l-1}
\sum\limits_{\alpha\in R}|\alpha|^2\left(E_1\left(\bfu_\alpha+\lan\kappa,\alpha\ran\tau+\frac{q}l\right)+2\pi
i\lan\kappa,\alpha\ran\right)\gt_\alpha^{q,a}(0)\gt_{-\alpha}^{-q,a}(0)\\
 \qquad{}
=\frac12\sum\limits_{q=0}^{l-1}
\sum\limits_{\alpha\in R}|\alpha|^2\left(E_1\left(\bfu_\alpha+\lan\kappa,\alpha\ran\tau+\frac{q}l\right)+2\pi
i\lan\kappa,\alpha\ran\right)\frac{p_\alpha}{\sqrt{l}}\exp\left(-2\pi i\frac{q}{l}\right)\gh^{0,a}_\alpha\\
\qquad{}
=l\sum\limits_{\alpha\in \tilde\Pi}\gh_{\alpha}^{0,a}(0)\p_{\bfu_{\alpha}}\left\{ \log
\prod\limits_{q=0}^{l-1}\prod\limits_{\alpha\in R}
\vartheta\left(\bfu_\alpha+\lan\kappa,\alpha\ran\tau
+\frac{q}l\right)^{\frac{p_\alpha|\alpha^2|}{2l\sqrt{l}}\exp\left(-2\pi i\frac{q}{l}\right)}\right\}.
\end{gather*}
The term
\[
\frac12\sum\limits_{q=0}^{l-1}
\sum\limits_{\alpha\in R}|\alpha|^2 2\pi i\lan\kappa,\alpha\ran\frac{p_\alpha}{\sqrt{l}}
\exp\left(-2\pi i\frac{q}{l}\right)\gh^{0,a}_\alpha
\]
vanishes because of summation over $q$.
Notice also that the obtained scalar expression does not depend on $\{z_c\}$.
Then, the equation (\ref{kze1}) gives
\begin{gather*}
\Bigg(\p_a+l \sum\limits_{\alpha\in \tilde\Pi}\gh_{\alpha}^{0,a}(0)\p_{\bfu_{\alpha}}+
\sum\limits_{q=0}^{l-1}\sum\limits_{c\neq a}
\Bigg(\sum\limits_{\alpha\in R}|\alpha|^2\varphi_\alpha^{q}(\bfu_\alpha,z_c-z_a)
\gt_\alpha^{q,c}(0)\gt_{-\alpha}^{-q,a}(0)\nonumber\\
\qquad{}+ \sum\limits_{\alpha\in\tilde\Pi}\varphi_\alpha^{q}(0,z_c-z_a)
\gH_\alpha^{q,c}(0)\gh_{-\alpha}^{-q,a}(0)\Bigg)\Bigg)\tilde F=0,
\end{gather*}
where
\[
{\tilde F}=F\prod\limits_{q=0}^{l-1}\prod\limits_{\alpha\in R}
\vartheta\left(\bfu_\alpha + \lan\kappa,\alpha\ran\tau + \frac{q}{l}\right)
^{-\frac{p_\alpha|\alpha^2|}{2l\sqrt{l}}\exp\left(-2\pi i\frac{q}{l}\right)}.
\]
This is the f\/irst set of equations in \eqref{t1}.
In order to obtain the second one (the KZB connec\-tion~$\nabla_\tau$ along $\tau$) one should
use (\ref{Vir3}).
It is needed to compute $L_{-2}^a$.
The later arises from the local expansion of~(\ref{mmfi}) for $k=1$.
Then the following identities should be used
\begin{gather*}
\p_z\phi(u,z)=\phi(u,z)\big(E_1(z+u)-E_1(z)\big)=f(u,z)+\big(E_1(u)-E_1(z)\big)\phi(u,z),
\end{gather*}
where $f(u,z)=\p_u\phi(u,z)$ for $\gt( - 2)\gt(0)$-terms and
\begin{gather*}
 \phi(u,z-z_a)\phi(-u,z-z_b)=-\phi(u,z-z_a)\phi(u,z_b-z)\nonumber\\
 \qquad{}
 =-\phi(u,z_b-z_a)\big(E_1(u)+E_1(z-z_a)+E_1(z_b-z)-E_1(u+z_b-z_a)\big)\nonumber\\
 \qquad{}
 =f(u,z_b-z_a)+\phi(u,z_b-z_a)\big(E_1(z-z_b)-E_1(z-z_a)\big)
\end{gather*}
for $\gt( - 1)\gt( - 1)$-terms.
On the other hand $\nabla_\tau$ is a unique f\/lat connection for given $\nabla_a$~(\ref{t2}).
The f\/inal answer is given below in Section~\ref{section4.6}.
This answer is verif\/ied in  Appendix~\ref{appendixC}.

\subsection[Classical $r$-matrix]{Classical $\boldsymbol{r}$-matrix}\label{section4.5}

The construction of the KZB connection is based on the
classical dynamical elliptic $r$-matrix def\/ined as sections of bundles over elliptic curves \cite{dirac1,int2,dirac2}.
For
trivial $G$-bundles our list coincides with the elliptic $r$-matrices were def\/ined in~\cite{EV}.
A more general class of elliptic  $r$-mat\-ri\-ces  was constructed
in \cite{ES1,ES2}.
The latter classif\/ication includes our list though it was derived from dif\/ferent
postulates.

\subsubsection[Axiomatic description of $r$-matrices]{Axiomatic description of $\boldsymbol{r}$-matrices}

The classical dynamical $r$-matrix is a meromorphic one form $r=r(\bfu,z)dz$, $(\bfu\in\tilde\gh_0)$
on $\mC$ taking values in
$\gg\otimes\gg$ that satisf\/ies the following conditions:
\begin{enumerate}\itemsep=0pt
\item $r(z)$ has a pole at $z=0$ and
\begin{gather*}
\operatorname{Res}\big|_{z=0}\;r(z)=C_2=\frac12\sum\limits_{k=0}^{l-1} \sum\limits_{\alpha\in R}
|\alpha|^2\gt^{k}_{\alpha} \otimes \gt^{-k}_{-\alpha}+ \sum\limits_{k=0}^{l-1}\sum_{\alpha\in\Pi}
\gH^{k}_{\alpha}\otimes\gh^{-k}_{\alpha},
\end{gather*}
where $\gt^{k}_{\alpha}$, $\gH^{k}_{\alpha}$, $\gh^{-k}_{\alpha}$ are generators of the GS basis in~$\gg$ (see
Appendix~\ref{appendixA}).
If $V$ is a $\gg$-module, then $C_2$ acts by the permutation on $V\otimes V$.

\item
Behavior under the shifts by the generators of the lattice $\mZ\oplus\tau\mZ$:
\begin{gather}\label{qpp}
r(z+1)=\operatorname{Ad}_\clQ r(z),\qquad r(z+\tau)=2\pi\imath\sum_{\alpha\in\Pi}
\gH^{0}_{\alpha}\otimes\gh^{0}_{\alpha}+\operatorname{Ad}_{\Lambda_j}r(z),
\end{gather}
where the $\operatorname{Ad}$-action is taken with respect to the f\/irst factor in $\gg\otimes\gg$.
Here $\clQ=\bfe(\kappa)$,
$\Lambda^{(j)}=\Lambda_0\bfe(\bfu)$ (see~(\ref{trm})), $\{\gh^{0}_{\alpha}\}$ ($\{\gH^{0}_{\alpha}\}$)
is the simple coroot
basis
(the dual basis) in the invariant subalgebra $\tilde\gg_0$ (see Appendix A).
It means that $r$ is a connection in the $\gg\otimes\gg$-bundle over $\Sigma_\tau$.

\item
\emph{The classical dynamical Yang--Baxter equation $($CDYBE$)$.}
It follows from~1
that $r(z)$ can be represented as
\[
r(z)=\frac12\sum\limits_{k=0}^{l-1}
\sum\limits_{\alpha\in R} \Phi^k_\alpha(z) |\alpha|^2\gt^{k}_{\alpha} \otimes \gt^{-k}_{-\alpha}+
\sum\limits_{k=0}^{l-1}\sum_{\alpha\in\Pi} \Psi^k_\alpha(z)\gH^{k}_{\alpha}\otimes\gh^{-k}_{\alpha}.
\]
Then $r(z)$ is a solution of CDYBE:
\begin{gather}
 \left[r_{12}(z_{12}),r_{13}(z_{13})\right]
+\left[r_{12}(z_{12}),r_{23}(z_{23})\right]
+\left[r_{13}(z_{13}),r_{23}(z_{23})\right]\nonumber\\
 \qquad
{} -\sqrt{l}\sum\limits_{k=0}^{l-1}\sum\limits_{\alpha\in R}\frac{|\alpha|^2}{2} \gt^{k}_{\alpha}\otimes
\gt^{-k}_{-\alpha}\otimes \bar{\gh}_{\alpha}^{0}\p_1\Phi_\alpha^k(\bfu,z-w)\nonumber\\
 \qquad
{}-\frac{|\alpha|^2}{2}\gt^{k}_{\alpha}\otimes
\bar{\gh}_{\alpha}^{0}\otimes t^{-k}_{-\alpha}\p_1\Phi_\alpha^k(\bfu,z-x)\nonumber\\
\qquad
{}+\frac{|\alpha|^2}{2}\bar{\gh}_{\alpha}^{0}\otimes
\gt^{k}_{\alpha}\otimes\gt^{-k}_{-\alpha}\p_1\Phi_\alpha^k(\bfu,w-x)=0,
\qquad  z_{ij}=z_i-z_j ,\label{dfg5}
\end{gather}
where $\p_1$ is the dif\/ferentiation with respect to the f\/irst argument.
\item
The unitarity
\begin{gather*}
r^{12}(\bfu,z)+r^{21}(\bfu,-z)=0.
\end{gather*}

\item
The zero weight condition
\begin{gather*}
\left[X\otimes 1+1\otimes X,r(\bfu,z)\right]=0, \qquad X\in\gh.
\end{gather*}
\end{enumerate}

\begin{lemma}
$\bullet$ Any $r'$-matrix satisfying $1$--$5$ has the form
\begin{gather*}
r'(\bfu,z)= r(\bfu,z)+ \delta r(\bfu),
\end{gather*}
where
\begin{gather}
r(\bfu,z)=r_{\gH}(\bfu,z)+ r_{R}(z),\label{prm1}\\
r_{R}(\bfu,z)=\frac12\sum\limits_{k=0}^{l-1} \sum\limits_{\alpha\in R}r^k_{\alpha}(\bfu,z),
\qquad r^k_{\alpha}(\bfu,z)=
|\alpha|^2\varphi^{k}_{\alpha}(\bfu,z) \gt^{k}_{\alpha} \otimes \gt^{-k}_{-\alpha},\nonumber\\ 
r_{\gH}(z)=
\sum\limits_{k=0}^{l-1}\sum_{\alpha\in\Pi}r^0_\alpha(z),\qquad r^0_\alpha(z)= \varphi^{k}_{0}(z)
\gH^{k}_{\alpha}\otimes\gh^{-k}_{\alpha},\nonumber
\end{gather}
satisf\/ies \emph{1--5} and
$\varphi^k_\beta(\bfx,z)$ is def\/ined
in 
\eqref{kfi}, $\varphi^{k}_{0}(z)=\phi(k/l,z) $, $\varphi^0_0(z)=E_1(z)$.

$\bullet$
$\delta r(\bfu)\in\tilde\gh_0\otimes\tilde\gh_0$
\begin{gather*}
\delta r(\bfu)=\sum_{\alpha,\beta\in\tilde\Pi}A_{\alpha\beta}\gH^{0}_{\alpha}\otimes\gh^{0}_{\beta},
\qquad A_{\alpha\beta}=-A_{\beta\alpha}.
\end{gather*}
$\bullet$
$\delta r(\bfu)$ is generated by the gauge transformation
\begin{gather*}
\delta r(\bfu)=-l\sum\limits_{\alpha\in\tilde\Pi}\big(\partial_{u_{\hat{\alpha}}}f\big) f^{-1}\otimes\gh^{0}_\alpha,
\qquad \big(\partial_{u_{\hat{\alpha}}}f\big) f^{-1}\in\tilde\gh_0, \qquad f=f(\bfu)\in \tilde H_0,
\end{gather*}
where $\tilde H_0$ is a Cartan subgroup of the invariant subgroup $\tilde G_0\subset G$
$($see Table~{\rm 1} in~{\rm \cite{LOSZ})}.
\end{lemma}

\begin{proof}
It follows from the properties of the functions $\varphi^{k}_{\alpha}(\bfu,z)$, $\varphi^{k}_{0}(\bfu,z)$ described
in the Appendix B that $r(\bfu,z)$ satisfy~1 and~2.
It was proved in \cite{LOSZ} that it is a solution of the CDYBE.
This
sum is a classical dynamical $r$-matrix corresponding to a non-trivial characteristic class def\/ined by~(\ref{qpp}).
The
conditions~4 and~5 can be checked as well.
The conditions 1--3, 5 def\/ine the $r$-matrix up to a constant
($z$-independent) Cartan term~$\delta r$.
Then it follows from~4 that $A_{\alpha\beta}$ is
antisymmetric.

Next we wish to prove that locally $A_{\alpha\beta}=-l(\partial_{u_{{\alpha}}}(f)f^{-1})_\beta$
for some $ f\in\tilde H_0$.
The
twisted r-matrix must satisfy the CDYB equation.
Plugging $r+\delta r$ into (\ref{dfg5}) we see that the ``commutator'' part
vanishes identically since $[r^{ab},\delta r^{ac}]+[r^{ab},\delta r^{bc}] \equiv 0$ due to
\[
\big[\gt_\alpha^{k,a},\gh^{0,a}_\beta\big]\otimes \gt^{-k,a}_{-\alpha}+\gt_\alpha^{k,a}\otimes
\big[\gt^{-k,a}_{-\alpha},\gh^{0,a}_\beta\big]=0.
\]
The ``derivative'' part of (\ref{dfg5}) yields
$\partial_{u_{\hat{\alpha}}}A_{\beta\gamma}+\partial_{u_{\hat{\gamma}}}A_{\alpha\beta}+
\partial_{u_{\hat{\beta}}}A_{\gamma\alpha}=0$ or
\begin{gather*}
dA=0,\qquad A=\sum\limits_{\alpha,\beta\in \Pi }A_{\alpha\beta}d u_{{\alpha}}\wedge d u_{{\beta}}\in\clM_G.
\end{gather*}
The term $\delta r$ is called the \emph{dynamical twist} of the $r$-matrix.
The statement follows from the Poincar\'e lemma.
\end{proof}

\subsubsection[$r$-matrices as sections of bundles over moduli spaces]{$\boldsymbol{r}$-matrices as sections of bundles over moduli spaces}

Consider the behavior of the $r$-matrix \eqref{prm1} under
the action of
latices $\tau \tilde Q^\vee\oplus \tilde Q^\vee$ (\ref{twsc}) and $\tau\tilde P^\vee\oplus \tilde P^\vee$ (\ref{twad})
on the dynamical parameter $\bfu$.
It follows from (\ref{A.300}), (\ref{kfi}) and (\ref{A.14}) that the $r$-matrices has distinct type
of quasi-periodicities with respect $\bfu\in\tilde\gh_0$.
Let $\beta^\vee\in\tilde\Pi^\vee$ be a simple coroot, corresponding to the invariant algebra
$\tilde\gg_0$.
For $\alpha\in R$ def\/ine the integers $n_{\alpha,\beta}=\lan\alpha,\beta^\vee\ran$.
Then we f\/ind
\[
r^k_\alpha\big(\bfu+\beta^\vee,z\big)=r^k_\alpha(\bfu,z),\qquad
r^k_\alpha\big(\bfu+\tau\beta^\vee,z\big)=\bfe\big({-}n_{\alpha,\beta}z\big)r^k_\alpha(\bfu,z).
\]
Let $\Xi^\vee$ be a basis of fundamental co-weights
dual to the basis $\Pi$, and $\tilde\varpi^\vee$ is a fundamental coweight in $\tilde P^\vee$.
Since $\tilde P^\vee$ is a sublattice of $P^\vee$, the
weight $\tilde\varpi^\vee$ can be decomposed in the basis of the fundamental co-weights
$\tilde\varpi^\vee=\sum_{\nu^\vee\in \Xi^\vee}n_{\nu}^{\varpi}\nu^\vee$, where $n_{\nu}^{\varpi}\in\mZ$.
As above we f\/ind
\[
r^k_\alpha(\bfu+\tilde\varpi^\vee,z)
=r^k_\alpha(\bfu,z),\qquad r^k_\alpha(\bfu+\tau\tilde\varpi^\vee,z)
=\bfe\big({-}n_{\nu}^{\varpi}\delta_{\lan\nu^\vee,\alpha\ran}z\big)r^k_\alpha(\bfu,z).
\]

On the other hand, due to the $\Lambda$-invariance of $\tilde Q^\vee$, we have
$\lan\beta^\vee,\lambda^m(\alpha)\ran=\lan\beta^\vee,\alpha\ran$.
Therefore,
$\operatorname{Ad}_{\exp2\pi\imath\beta^\vee}E_{\lambda^m(\alpha)}=\bfe(\lan\alpha,\beta^\vee\ran)E_{\lambda^m(\alpha)}$.
Then from (\ref{ft}) we f\/ind that
\[
\operatorname{Ad}_{\exp(-2\pi\imath\beta^\vee z)}\gt_\alpha^a=\bfe(-n_{\alpha,\beta}z)\gt_\alpha^a.
\]
Similarly, due to $\Lambda$-invariance of $\tilde P^\vee$, we have also
\[
\operatorname{Ad}_{\exp(-2\pi\imath\tilde\varpi^\vee z)}\gt_\alpha^a
=\bfe\big({-}n_{\nu}^{\varpi}\delta_{\lan\nu^\vee,\alpha\ran}z\big)\gt_\alpha^a.
\]

Since the Cartan part $r_\gH$ of the $r$-matrix does not depend on $\bfu$ we come to the relations
\begin{alignat}{3}
& r\big(\bfu+\beta^\vee,z\big)=r(\bfu,z),\qquad &&
r\big(\bfu+\tau\beta^\vee,z\big)= \operatorname{Ad}_{\exp(-2\pi\imath\beta^\vee z)}r(\bfu,z),& \label{qpm}\\
& r\big(\bfu+\tilde\varpi^\vee,z\big)=r(\bfu,z),\qquad &&
r\big(\bfu+\tau\tilde\varpi^\vee,z\big)=\operatorname{Ad}_{\exp(-2\pi\imath\tilde\varpi^\vee z)}r(\bfu,z).& \label{qpm1}
\end{alignat}
In all cases the adjoint actions $\operatorname{Ad}_h$ act on the f\/irst component of the tensor product and play the role
of the clutching operators.

Let $x(k,\alpha)=\lan\bfu,\alpha+\kappa\tau\ran+k/l$. Then $r(\bfu,z)$ is singular when $x(k,\alpha)\to 0$
(see~(\ref{mmfi0}) and~(\ref{A.300}))
\begin{gather}\label{sms}
r(\bfu,z)=|\alpha|^2\bfe(\lan\kappa,\alpha\ran z)\left(\frac1{x(k,\alpha)}+O(1)\right)\gt^{k}_{\alpha}
\otimes \gt^{-k}_{-\alpha}.
\end{gather}
It means that $ r(\bfu,z)$ are sections of the bundles over the moduli spaces $C^{\rm sc}_j$~(\ref{tsc}),
or~$C^{\rm ad}_j$~(\ref{tad}) with sections taking values in $\gg\otimes\gg$
with the quasi-periodicities (\ref{qpm}), (\ref{qpm1}) and with the singularities~(\ref{sms}).

\subsection{KZB connection related to elliptic curves}\label{section4.6}

As it was established the part of connection related to the moving points coincides with the introduced above $r$-matrix.
Here we prove that this connection is f\/lat.
Consider the following dif\/ferential operators
\begin{gather}
\nabla_a=\p_{z_a}+\hat{\p}^a+\sum\limits_{c\neq a}r^{ac},\label{t2}\\
\nabla_\tau=2 \pi i \p_\tau+\Delta+\frac12\sum\limits_{b,d}f^{bd},\nonumber 
\end{gather}
with
\begin{gather*}
r^{ac}=\sum\limits_{k=0}^{l-1} \sum\limits_{\alpha \in R}|\alpha|^2
\varphi^{k}_{\alpha}(\bfu,z_a-z_c) \gt^{k,a}_{\alpha} {\gt^{-k, c}_{-\alpha}} +
\sum\limits_{k=0}^{l-1}\sum\limits_{\alpha\in \Pi} \varphi_{0}^{k}(\bfu,z_a-z_c)
\gH_{\alpha}^{k,a}\gh_{\alpha}^{k,c},\\ 
f^{ac}=\sum\limits_{k=0}^{l-1} \sum\limits_{\alpha \in R}
|\alpha|^2 f^{k}_{\alpha}(\bfu,z_a-z_c)
\gt^{k,a}_{\alpha} {\gt^{-k, c}_{-\alpha}} +
\sum\limits_{k=0}^{l-1}\sum\limits_{\alpha\in \Pi} f_{0}^{k}(\bfu,z_a-z_c) \gH_{\alpha}^{k,a}
\gh_{\alpha}^{k,c},
\end{gather*}
where $\gt^{k,a}_{\alpha}=1\otimes\cdots\otimes 1\otimes \gt^{k}_{\alpha}\otimes 1\otimes\cdots\otimes 1$
(with $\gt^{k}_{\alpha}$ on the $a$-th
place) and similarly
for the generators~$\gH^{k,a}_{\alpha}$ and~$\gh^{k,a}_{\alpha}$.\footnote{For brevity we write $\gt^{k,a}_{\alpha}$, $\gh^{k,a}_{\alpha}$
instead of representations of these generators
in the spaces $V_{\mu_a}$.}
The following short notations are used here
\[
\hat{\p}^a=l \sum\limits_{\alpha \in \Pi } \gh^{0,a}_{\alpha} \partial_{\hat{\alpha}},\qquad
\Delta=\frac{l}{2}\sum\limits_{\alpha\in \Pi}
\sum\limits_{s=0}^{l-1} \p_{u_\alpha}\p_{u_{\lambda^s\hat{ \alpha}}}
\]
and
\begin{gather}
 \varphi^{k}_{\alpha}(\bfu,z)=e^{2\pi i\lan\kappa,\alpha\ran z}\phi
\left(\lan\bfu+\kappa \tau, \alpha\ran+\frac{k}{l},z\right),\nonumber\\
 f^{k}_{\alpha}(\bfu,z)=e^{2\pi i\lan\kappa,\alpha\ran z}f
\left(\lan\bfu+\kappa\tau,\alpha\ran+\frac{k}{l},z\right).\label{t303}
\end{gather}
From the def\/inition it follows that $r^{ac}=-r^{ca}$ and $f^{ac}=f^{ca}$.
Following (\ref{A3b}) and (\ref{mmfi0}) we put
\begin{gather}
\varphi^{0}_{0}(z)=E_{1}(z),\label{t305}\\
f^{0}_{0}(z)=\rho(z)=\frac12\big( E_{1}^{2}(z)-\wp(z)\big).\label{t306}
\end{gather}
Notice that
\[
f^{k}_{\alpha}(\bfu,0)
=-E_{2}\left(\langle u+\kappa\tau,\alpha\rangle +\frac{k}{l}\right)
=-\wp\left(\langle u+\kappa\tau,\alpha\rangle +\frac{k}{l}\right)-2\eta_1
\]
and, therefore
\begin{gather*}
f^{cc}=-\sum\limits_{k=0}^{l-1}\sum\limits_{\alpha\in R} |\alpha|^2
\wp^{k}_{\alpha} \gt^{k,c}_{\alpha} \gt^{-k,c}_{-\alpha} -\sum\limits_{k=0}^{l-1}
\sum\limits_{\alpha\in \Pi}
\gH^{k,c}_{\alpha} \gh^{-k,c}_{\alpha}-2l\eta_{1} C_{2}^{c},
\end{gather*}
where $C_{2}^{c}$ is the Casimir operator acting on the $c$-th component.
Recall that we study the following system of dif\/ferential equations
\begin{gather}\label{t1}
\nabla_aF=0,\quad a=1,\dots, n,\qquad
\nabla_\tau F=0.
\end{gather}
There are two types of the compatibility conditions of KZB equations (\ref{t1})
\begin{gather}\label{t5}
[\nabla_a,\nabla_b]F=0,\quad                   a,b=1,\dots, n,\qquad
\left[\nabla_a,\nabla_{\tau}\right]F=0,\quad  a=1,\dots, n.
\end{gather}
It is important to mention that the solutions of (\ref{t1}) $F$ are assumed to satisfy the following condition
\begin{gather}\label{t4}
\left( \sum\limits_{c=1}^n \gh^{0,c}_{\alpha}\right)F=0,\qquad \text{for any} \ \ \alpha\in \tilde\Pi.
\end{gather}

\begin{proposition}\label{predl1}
The upper equations in \eqref{t5} $[\nabla_a,\nabla_b]=0$ are valid for the $r$-matrix \eqref{prm1}
on the space of solutions
of \eqref{t1} satisfying \eqref{t4}.
They follow from the classical dynamical Yang--Baxter equations
\begin{gather}\label{t501}
\big[r^{ab},r^{ac}\big]+\big[r^{ab},r^{bc}\big]+\big[r^{ac},r^{bc}\big]+ \big[\hat{\p}^a,r^{bc}\big]
+\big[\hat{\p}^c,r^{ab}\big]+\big[\hat{\p}^b,r^{ca}\big]=0.
\end{gather}
\end{proposition}

\begin{proposition}\label{predl2}
The lower equations in \eqref{t5} $\big[\nabla_a,\nabla_{\tau}\big]=0$ are valid for the $r$-matrix \eqref{t303}
on the space of solutions of~\eqref{t1} satisfying~\eqref{t4}.
\end{proposition}
The proofs of these statements are given in the Appendix~\ref{appendixC}.

Let us also remark that the non-trivial trigonometric and rational limits of the above formulae can be obtained via
procedures described in \cite{trig3,trig2, trig1}.

\appendix

\section{Generalized Sine (GS) basis in simple Lie algebras}\label{appendixA}

Let $\clZ $ be a subgroup of the center $\clZ(\bar G)$ of $\bar{G}$, and consider a quotient group $G=\bar{G}/\clZ$.
Assume for
simplicity that $\clZ\big(\bar{G}\big)$ is cyclic.
The case $\operatorname{Spin}(4n)$ where $\clZ(G)=\mu_2\times\mu_2$ can be treated similarly.

Let us take an element $\zeta\in\clZ(\bar G)$ of order $l$, generating $\clZ$.
It def\/ines uniquely an element $\Lambda_0$ from
the Weyl group $W$ (see \cite{Bou, LOSZ}).
It is a symmetry of the corresponding extended Dynkin diagram and $(\Lambda_0)^l=\operatorname{Id}$.
$\Lambda_0$ generates a cyclic
group $\mu_l=\left(\Lambda_0,(\Lambda_0)^2,\ldots,(\Lambda_0)^l=1\right)$ isomorphic to a subgroup of $\clZ(\bar G)$.
Note that $l$ is a divisor
of $\operatorname{ord}(\clZ(\bar G))$.
Consider the action of $\Lambda_0$ on $\gg$.
Since $(\Lambda_0)^l=\operatorname{Id}$ we have a $l$-periodic gradation
\begin{gather}
\gg=\oplus_{a=0}^{l-1}\gg_a,
\qquad
\lambda(\gg_a)=\omega^a\gg_a,
\qquad
\omega=\exp\frac{2\pi i}l,
\qquad
\lambda=\operatorname{Ad}_{\Lambda_0},\label{gra}\\
[\gg_a,\gg_b]=\gg_{a+b} \quad  \operatorname{mod} l,\nonumber 
\end{gather}
where $\gg_0$ is a
subalgebra $\gg_0\subset\gg$ and the subspaces $\gg_a$ are its representations.
Since $\clQ$ and $\Lambda$ commute in the adjoint representations the root subspaces~$\gg_a$
are their common eigenspaces.

\textbf{GS-basis.}
Here we shortly reproduce the construction of the GS-basis following~\cite{LOSZ}.
Since $\Lambda_0\in W$ it preserves the root system $R$.
Def\/ine the quotient set $\clT_l=R/\mu_l$.
Then $R$ is represented as a~union of $\mu_l$-orbits $R=\cup_{\clT_l}\clO$.
We denote by $\clO(\babe)$ an orbit starting
from the root~$\beta$
\[
\clO(\babe)=\big\{\beta,\lambda(\beta),\ldots,\lambda^{l-1}(\beta)\big\},
\qquad
\babe\in \clT_l.
\]
The number of elements in an orbit $\clO$ (the length of $\clO$) is $l/p_\alpha=l_\alpha$,
where $p_\alpha$ is a divisor of $l$.
Let
$\nu_\alpha$ be a number of orbits $\clO_{\baal}$ of the length $l_\alpha$.
Then $\sharp R=\sum\nu_\alpha l_\alpha$.
Notice that if
$\clO(\babe)$ has length $l_\beta (l_\beta\neq 1)$, then the elements $\lambda^k\beta$
and $\lambda^{k+l_\beta}\beta$ coincide.
First, transform the root basis $\clE=\{E_\beta$, $\beta\in R\}$ in $\gL$.
Def\/ine an orbit in $\clE$
\[
\clE_{\beta}=\big\{E_\beta,E_{\lambda(\beta)},\ldots,E_{\lambda^{l-1}(\beta)}\big\}
\]
corresponding to $\clO(\babe)$.
Again
$\clE=\cup_{\babe\in\clT_l}\clE_{\babe}$.
For $\clO(\babe)$ def\/ine the set of integers
\begin{gather}\label{dc}
J_{p_\alpha}=\big\{a=mp_\alpha\;\big|\;m\in\mZ, \; a\text{ is def\/ined}\operatorname{mod}l\big\},
\qquad
p_\alpha=\frac{l}{l_{\alpha}}.
\end{gather}
Let $E_\alpha$ ($\alpha\in R$) be the root basis of $\gg$.
``The Fourier transform'' of the root basis on the orbit~$\clO(\babe)$ is def\/ined as
\begin{gather}\label{ft}
\gt^a_{\beta}=\frac1{\sqrt{l}}\sum_{m=0}^{l-1}\omega^{ma}E_{\lambda^m(\beta)},
\qquad
\omega=\exp\frac{2\pi i}{l},\qquad a\in J_\beta.
\end{gather}

Almost the same construction exists in $\gH$.
Again let $\Lambda_0$ generates the group $\mu_l$.
Since $\Lambda_0$ preserves the
extended Dynkin diagram, its action preserves the extended coroot system
$\Pi^{\vee {\rm ext}}=\Pi^\vee\cup \alpha^\vee_0$ in $\gH$.
Consider the quotient $\clK_l=\Pi^{\vee {\rm ext}}/\mu_l$.
Def\/ine an orbit $\clH(\baal)$ of length $l_\alpha=l/p_\alpha$ in $\Pi^{\vee {\rm ext}}$
passing through $H_\alpha\in\Pi^{\vee {\rm ext}}$
\[
\clH(\baal)=\big\{H_\alpha,H_{\lambda(\alpha)},\ldots,H_{\lambda^{l-1}(\alpha)}\big\},
\qquad
\baal\in\clK_l=\Pi^{\vee {\rm ext}}/\mu_l.
\]
The set $\Pi^{\vee {\rm ext}}$ is a union of $\clH(\baal)$:
\[
(\Pi^{\vee})^{{\rm ext}}=\cup_{\baal\in\clK_l}\clH(\baal).
\]
Def\/ine ``the Fourier transform''
\begin{gather*}
\gh^c_{\baal}=\frac1{\sqrt{l}}
\sum_{m=0}^{l-1}\omega^{mc}H_{\lambda^m(\alpha)},
\qquad
\omega=\exp\frac{2\pi i}{l},
\qquad
c\in J_\alpha\quad(\text{see \eqref{dc}}).
\end{gather*}

The basis $\gh^c_{\alpha}$ $(c\in J_\alpha$, $\baal\in\clK_l)$
is over-complete in $\gH$.
Namely, let $\clH(\baal_0)$ be an orbit passing through the minimal coroot
$\big\{H_{\alpha_0},H_{\lambda(\alpha_0)},\ldots,H_{\lambda^{l-1}(\alpha_0)}\big\}$.
Then the element $\gh^0_{\bar{\alpha}_0}$ is a linear combination of
elements $\gh^0_{-\baal}$, $(\alpha\in\Pi)$ and we should exclude it from the basis.
We replace the basis $\Pi^\vee$ in $\gH$
by
\begin{gather}\label{cb}
\gh^c_{\baal},\quad c\in J_\alpha,\qquad
\begin{cases}
\alpha\in\tilde\clK_l=\clK_l\setminus \clH(\baal_0),  & c=0, \\
\baal\in \clK_l, & c\neq 0.
\end{cases}
\end{gather}
As before there is a one-to-one map $\Pi^\vee\leftrightarrow\{\gh^c_{\baal}\}$.
The elements $(\gh^a_{\baal},\gt^a_{\baal})$
form GS basis in $\gg_{(l-a)}$~(\ref{gra}).
The dual basis is generated by elements~$\gH^a_{\baal}$
\begin{gather}\label{dbh}
\big(\gH^a_{\baal}, \gh^b_{\babe}\big)=\delta^{(a+b,0(\operatorname{mod} l))}\delta_{\alpha,\beta},
\qquad
\gH^a_{\baal}=\sum_{\beta\in\Pi}(\clA^a_{\alpha,\beta})^{-1}\gh^{-a}_{\babe},
\qquad
\gh^{a}_{\babe}=\sum_{\alpha\in\Pi}(\clA^{-a}_{\alpha,\beta})\gH^{-a}_{\baal},
\end{gather}
where
\[
\clA^a_{\alpha,\beta}=\frac{2}{(\beta,\beta)}\sum _{s=0}^{l-1}\omega^{-sa} a_{\beta,\lambda^s(\alpha)}
\]
and $a_{\alpha,\beta}$ is the Cartan matrix of~$\gg$.

The $\lambda$-invariant subalgebra $\gg_0$ contains the subspace
\begin{gather*}
V=\Bigg\{\sum_{\babe\in \clT'_l}a_{\babe}\gt^0_{\babe},\  a_{\babe}\in\mC\Bigg\}.
\end{gather*}
Then $\gg_0$ is a sum of $\tilde{\gg}_0$ and $V$
\begin{gather*}
\gg_0=\tilde{\gg}_0 \oplus V.
\end{gather*}
In the invariant simple algebra $\tilde{\gg}_0$ instead of
the basis $(\gh^0_{\baal},\gt^0_{\babe})$ we can use the Chevalley basis
and incorporate it in the GS-basis
\begin{gather*}
\big\{\gh^0_{\baal},\gt^0_{\babe}\big\}\to
\big\{\tilde\gg_0=\big(H_{\tilde\alpha},\tilde\alpha\in\tilde\Pi,\;
E_{\tilde\beta},\tilde\beta\in\tilde R\big),\;
V=\big(\gt^0_{\babe},\babe\in\clT'\big)\big\},
\end{gather*}
where $\tilde\Pi$ is a system of simple roots constructed by the averaging of the $\lambda$ action on $\Pi^{{\rm ext}}$,
and $\tilde R$ is a
system of roots of $\tilde{\gg}_0$ generated by $\tilde\Pi$.
We have the following action of the adjoint operators on the GS basis:
\begin{gather}\label{qc1}
\operatorname{Ad}_{\Lambda}\big(\gt^c_{\babe}\big)
=\bfe\left(\lan\tilde\bfu,\beta\ran-\frac{c}{l}\right)\gt^c_{\babe},
\qquad
 \operatorname{Ad}_{\Lambda}\big(\gh^c_{\babe}\big)=\bfe\left(-\frac{c}{l}\right)\gh^c_{\babe},
\qquad
\bfe(x)=\exp(2\pi ix).
\end{gather}
In addition,
\begin{gather}
\operatorname{Ad}_{\clQ}\big(\gh^c_{\babe}\big)
=\gh^c_{\babe},\qquad \operatorname{Ad}_{\clQ} (H_{\tilde\alpha})=H_{\tilde\alpha},\label{qc}\\
\operatorname{Ad}_{\clQ}\big(\gt^c_{\babe}\big)=\bfe(\lan\kappa,\beta\ran)\gt^c_{\babe},
\qquad \operatorname{Ad}_{\clQ}(E_{\tilde\alpha})=\bfe\lan\kappa,\tilde\alpha\ran E_{\tilde\alpha}.\label{qr}
\end{gather}
There are also the evident relations
\[
\operatorname{Ad}_{\Lambda}( E_{\tilde\alpha})=\bfe(\lan\tilde\bfu,\tilde\alpha\ran)E_{\tilde\alpha},
\qquad
\operatorname{Ad}_{\Lambda}(H_{\tilde\alpha})=H_{\tilde\alpha},\qquad \tilde\bfu\in\tilde\gh.
\]
In particular,
\begin{gather*}
\operatorname{Ad}_{\Lambda}\big(\gt^c_{\babe}\big)=\bfe\left(\lan\tilde\bfu,\beta\ran-\frac{c}{l}\right)\gt^c_{\babe},
\qquad
\operatorname{Ad}_{\Lambda}\big(\gh^c_{\babe}\big)=\bfe\left(-\frac{c}{l}\right)\gh^c_{\babe},
\qquad
\bfe(x)=\exp(2\pi ix).
\end{gather*}

\emph{Commutation relations in the GS basis:}
\begin{gather*}
\big[\gt^{a}_{\alpha},\gt^{b}_{\beta}\big]=
\begin{cases}
\displaystyle \frac{1}{\sqrt{l}}\sum\limits_{s=0}^{l-1} \omega^{bs}
C_{\alpha,\lambda^s\beta}\gt^{a+b}_{\alpha+\lambda^s\beta},\quad &\alpha\neq-\lambda^{s}\beta,\\
\displaystyle \frac{p_{\alpha}}{\sqrt{l}}\omega^{sb}\gh^{a+b}_{\alpha} ,  \quad &\alpha=-\lambda^{s}\beta,
\end{cases}
\\
\big[\gh^{k}_{\alpha},\gt^{m}_{\beta}\big]
=\frac{1}{\sqrt{l}}\sum\limits_{s=0}^{l-1}\omega^{-ks}\frac{2(\alpha,\lambda^{s}\beta)}{(\alpha,\alpha)}
\gt^{k+m}_{\beta},\nonumber\\
 \big[\gH^{k}_{\alpha},\gt^{m}_{\beta}\big]
=\frac{1}{\sqrt{l}}\sum\limits_{s=0}^{l-1}\omega^{- ks }\frac{(\alpha,\alpha)}{2}({\hat\alpha},\lambda^{s}\beta)
\gt^{k+m}_{\beta}.
\end{gather*}

\section{Elliptic functions}\label{appendixB}

The basic function is the theta-function
\begin{gather*}
\vartheta(z|\tau)=q^{\frac {1}{8}}\sum_{n\in {\bf Z}}( - 1)^ne^{\pi i(n(n+1)\tau+2nz)}.
\end{gather*}
It is a holomorphic function on $\mC$ with simple zeroes at the lattice $\tau\mZ+\mZ$ and the quasi-periodicities
\begin{gather*}
\vartheta(z+1)=-\vartheta(z),\qquad \vartheta(z+\tau)=-q^{-\frac12}e^{-2\pi iz}\vartheta(z).
\end{gather*}
Def\/ine the ration of the theta-functions
\begin{gather}\label{phi}
\phi(u,z)=\frac{\vartheta(u+z)\vartheta'(0)}{\vartheta(u)\vartheta(z)}.
\end{gather}
Then
\begin{gather}\label{A.300}
\phi(u,z)=\phi(z,u),\qquad \phi(-u,-z)=-\phi(u,z).
\end{gather}

\emph{Related functions:}
\begin{gather}
\varphi^m_\beta(\bfu,z)
=\bfe(\lan\kappa,\beta\ran z)\phi\left(\lan\bfu+\kappa\tau,\beta\ran+\frac{m}{l},z\right),\label{kfi}\\
\varphi_\beta^{m,k}(\bfu,z)
=\p^k_z\Bigl(\bfe(\lan\kappa,\beta\ran z)\phi\left(\lan \bfu+\kappa\tau,\beta\ran+\frac{m}{l},z\right)\Bigr),
\qquad \varphi_\beta^{m,0}=\varphi_\beta^{m} ,\label{mmfi}\\
f(u,z)=\p_u\phi(u,z),\label{A3c}\\
f(u,z)=\phi(u,z)(E_1(u+z)-E_1(u)).\label{A3b}
\end{gather}
$\phi(u,z)$ has a pole at $z=0$ and
\begin{gather}\label{mmfi0}
\phi(u,z)=\frac{1}{z}+E_1(u)+\frac{z}{2}\big(E_1^2(u)-\wp(u)\big)+\cdots.
\end{gather}
Similarly,
\begin{gather}\label{mmfi2}
\varphi_\beta^{m}(\bfu,z)
=\frac{1}{z}+E_1\left(\lan \bfu+\kappa\tau,\beta\ran+\frac{m}{l}\right)
+2\pi\imath\lan\kappa,\beta\ran+\frac{z}{2}\big(E_1^2(u)-\wp(u)\big)+\cdots,
\end{gather}
where $E_1$ is (\ref{A.1}). It follows from this expansion that
\begin{gather}
 \varphi_\beta^{m,1}(\bfu,z)=-\frac1{z^{2}}+\frac12\big(E_1^2(u)-\wp(u)\big)+\cdots,\nonumber\\
 \dots\dots\dots\dots\dots\dots\dots\dots\dots\dots\dots\dots\dots\dots\label{mmfi1}\\
 \varphi_\beta^{m,k}(\bfu,z)=\frac{( - 1)^{k}}{z^{k+1}}+c(m,k)+\cdots. \nonumber
\end{gather}
In other words $\varphi_\beta^{m,k}(\bfu,z)$ has not poles of order less than $k+1$.

{\it The Eisenstein functions:}
\begin{gather}\label{A.1}
E_1(z|\tau)=\p_z\log\vartheta(z|\tau), \qquad E_1(z|\tau)\sim\frac1{z}-2\eta_1z+\cdots,
\end{gather}
where
\begin{gather}
\eta_1(\tau)=\frac{3}{\pi^2} \sum_{m=-\infty}^{\infty} \sum_{n=-\infty}^{\infty '}
\frac{1}{(m\tau+n)^2}=\frac{24}{2\pi i}\frac{\eta'(\tau)}{\eta(\tau)},
\qquad
\eta(\tau)=q^{\frac{1}{24}}\prod_{n>0}\big(1-q^n\big),\nonumber\\
E_2(z|\tau)=-\p_zE_1(z|\tau)=\p_z^2\log\vartheta(z|\tau), \qquad E_2(z|\tau)\sim\frac1{z^2}+2\eta_1,\label{A.2}
\end{gather}
and more general for $k>2$
\begin{gather}\label{A.29}
E_k(z|\tau)= (-\p_z)^{k+1}\log\vartheta(z|\tau), \qquad E_k(z|\tau)\sim\frac1{z^{k}}+\cdots.
\end{gather}

{\it Relation to the Weierstrass functions:}
\begin{gather*}
\zeta(z,\tau)=E_1(z,\tau)+2\eta_1(\tau)z,\qquad 
\wp(z,\tau)=E_2(z,\tau)-2\eta_1(\tau).
\end{gather*}

{\it Quasi-periodicity:}
\begin{gather}
\vartheta(z+1)=-\vartheta(z),\qquad \vartheta(z+\tau)=-q^{-\frac12}e^{-2\pi iz}\vartheta(z),\label{A.11}\\
E_1(z+1)=E_1(z),\qquad E_1(z+\tau)=E_1(z)-2\pi i,\label{A.12}\\
E_k(z+1)=E_k(z),\qquad E_k(z+\tau)=E_k(z),\qquad k>1,\label{A.13}\\
\phi(u,z+1)=\phi(u,z),\qquad \phi(u,z+\tau)=e^{-2\pi \imath u}\phi(u,z),\label{A.14}\\
\varphi_\beta^{m,k}(\bfu,z + 1) = \bfe(\lan\kappa,\beta\ran)\varphi_\beta^{m,k}(\bfu,z),
\nonumber\\
\varphi_\beta^{m,k}(\bfu,z + \tau) = \bfe\left( - \lan\bfu,\beta\ran - \frac{m}{l}\right)
\varphi_\beta^{m,k}(\bfu,z),\label{A.14a}\\
f(u,z+1)=f(u,z),\qquad f(u,z+\tau)=e^{-2\pi \imath u}f(u,z)-2\pi\imath\phi(u,z).\nonumber
\end{gather}

The following identities are also used here
\begin{gather*}
2\pi i \p_{\tau} \phi(u,z)=\p_{z}\p_{u} \phi(u,z)=\p_{z} f(u,z)
\end{gather*}
and for the functions \eqref{t303} this identity takes the form
\begin{gather}\label{ap212}
2\pi i \p_{\tau} \varphi^{m}_{\alpha}(z)=\p_{z} f^{k}_{\alpha}(z).
\end{gather}

\emph{Fay identity:}
\begin{gather*}
\phi(u_1,z_1)\phi(u_2,z_2)-\phi(u_1 + u_2,z_1)\phi(u_2,z_2 - z_1)-\phi(u_1+u_2,z_2)\phi(u_1,z_1 - z_2)=0.
\end{gather*}
Dif\/ferentiating over $u_2$ we f\/ind
\begin{gather*}
 \phi(u_1,z_1) f(u_2,z_2)-\phi(u_1+u_2,z_1)f(u_2,z_2-z_1)\nonumber\\
 \qquad{}=\phi(u_2,z_2-z_1)f(u_1+u_2,z_1)+\phi(u_1,z_1-z_2)f(u_1+u_2,z_2).
\end{gather*}
Substituting here
\begin{gather*}
u_1=\langle u+\kappa\tau,\alpha+\beta\rangle +\frac{k+m}{l},\qquad
u_2=-\langle u+\kappa\tau,\beta\rangle -\frac{m}{l},\nonumber\\
z_1=z_a-z_c=z_{ac},\qquad
z_2=z_b-z_c=z_{bc},
\end{gather*}
and multiplying by appropriate exponential factor we can rewrite it in the form
\begin{gather}\label{ap33}
\varphi^{k}_{\alpha}(z_{ac}) f^{m}_{\beta}(z_{ab})-\varphi^{m}_{\beta} (z_{ab})f^{k}_{\alpha}(z_{ac})
+\varphi^{k+m}_{\alpha+\beta}(z_{ab})f^{k}_{\alpha}(z_{cb})- \varphi^{k+m}_{\alpha+\beta}(z_{ac})
f^{-m}_{-\beta}(z_{bc})=0.
\end{gather}
Taking the limit $m=0$, $\beta=0$ and using the expansion
\begin{gather*}
\phi(z,u)\sim\frac{1}{u}+E_{1}(z)+u\rho(z)+\cdots,
\end{gather*}
we f\/ind
\begin{gather}\label{ap215}
\varphi^{k}_{\alpha}(z_{ac}) \rho(z_{ab})-E_{1}(z_{ab}) f^{k}_{\alpha}(z_{ac})+ \varphi^{k}_{\alpha}(z_{ab})
f^{k}_{\alpha}(z_{bc})-\varphi^{k}_{\alpha}(z_{ac})\rho(z_{cb}) =\frac12\p_{u}f^{k}_{\alpha}(z_{ac}).
\end{gather}

\emph{More Fay identities:}
\begin{gather}
\varphi^{k}_{\alpha}(z_{ac})f^{m}_{\beta}(z_{ac})-\varphi^{m}_{\beta}(z_{ac}) f^{k}_{\alpha}(z_{ac})=
\varphi^{k+m}_{\alpha+\beta}(z_{ac})\big(\wp^{k}_{\alpha}-\wp^{m}_{\beta}\big),\label{ap29}\\
\varphi^{m}_{\beta}(z_{ac})f^{-m}_{-\beta}(z_{ac})-
\varphi^{-m}_{-\beta}(z_{ac})f^{m}_{\beta}(z_{ac})={E_{2}^{\prime}}^{m}_{\beta},\label{ap210}\\
\varphi^{k}_{\beta}(z_{ac})\wp^{k}_{\beta}-\varphi^{k}_{\beta}(z_{ac})\rho(z_{ac})+ E_{1}(z_{ac})f^{k}_{\beta}(z_{ac})=
\frac12 \p_{u}f^{k}_{\beta}(z_{ac}).\label{ap211}
\end{gather}
The last one follows from
\begin{gather*}
\p_{u}\phi(u,z)=\phi(u,z)\big(E_{1}(z+u)-E_{1}(u)\big)
\end{gather*}
and
\begin{gather*}
\big(E_{1}(z+u)-E_{1}(u)-E_{1}(z)\big)^2=\wp(z)+\wp(u)+\wp(z+u).
\end{gather*}

\section{Proofs of Propositions \ref{predl1} and \ref{predl2}}\label{appendixC}

\begin{proof}[Proof of Proposition \ref{predl1}.]
\begin{gather}\label{t6}
[\nabla_a,\nabla_b]=\big[\p_{z_a},r^{ba}\big]+\big[\hat{\p}^a,r^{ba}\big]\\
\phantom{\big[\nabla_a,\nabla_b\big]=}{}
+ \sum\limits_{c\neq a,b}\big[\hat{\p}^a,r^{bc}\big]-\big[\p_{z_b},r^{ab}\big]-\big[\hat{\p}^b,r^{ab}\big]
- \sum\limits_{c\neq a,b}\big[\hat{\p}^b,r^{ac}\big]+\sum\limits_{d\neq b}\sum\limits_{a\neq c}\big[r^{ac},r^{bd}\big].\nonumber
\end{gather}
First, notice that
\[
\big[\p_{z_a}+\p_{z_b},r^{ab}\big]=0.
\]
Secondly,
\[
\big[\hat{\p}^a,r^{ba}\big]-\big[\hat{\p}^b,r^{ab}\big]=-\big[\hat{\p}^a+\hat{\p}^b,r^{ab}\big]
=\sum\limits_{c\neq a,b}\big[\hat{\p}^c,r^{ab}\big]-\Bigg[\sum\limits_{c}\hat{\p}^c,r^{ab}\Bigg].
\]
Thirdly,
\[
\sum\limits_{d\neq b}\sum\limits_{a\neq c}\big[r^{ac},r^{bd}\big]=\sum\limits_{c\neq
a,b}\big(\big[r^{ac},r^{ba}\big]+\big[r^{ab},r^{bc}\big]+\big[r^{ac},r^{bc}\big]\big).
\]
Therefore,
\begin{gather}\label{t7}
\big[\nabla_a,\nabla_b\big]=\sum\limits_{c\neq a,b}\text{CDYB}^{abc}-\Bigg[\sum\limits_{c}\hat{\p}^c,r^{ab}\Bigg],
\end{gather}
where
\begin{gather}\label{t8}
\text{CDYB}^{abc}\!=\big[r^{ab},r^{ac}\big] + \big[r^{ab},r^{bc}\big] + \big[r^{ac},r^{bc}\big] +
\big[\hat{\p}^a,r^{bc}\big] + \big[\hat{\p}^c,r^{ab}\big] + \big[\hat{\p}^b,r^{ca}\big]\overset{\eqref{t501}}{=}0.\!\!\!
\end{gather}
and $\big[\sum\limits_{c}\hat{\p}^c,r^{ab}\big]\overset{\eqref{t4}}{=}0.$
\end{proof}

\begin{proof}[Proof of Proposition \ref{predl2}.] \quad

1.
\[
\big[\p_{z_a}f^{ac}\big]-2\pi i\big[\p_\tau r^{ac}\big]\overset{\eqref{ap212}}{=}0.
\]

2.
\begin{gather*}
\Bigg[\Delta, \sum\limits_{c\neq a} r^{a c}\Bigg]=
\frac{l}{2} \sum\limits_{c\neq a}\,\sum\limits_{\beta\in R} \sum\limits_{m=0}^{l-1}
|\beta|^2\p_{u} f^{m}_{\beta}(z_a - z_c)\big[\gh_{\beta}^{0,a},\gt_{\beta}^{m,a}\big]_+\gt^{-m,c}_{-\beta}\nonumber\\
\hphantom{\Bigg[\Delta, \sum\limits_{c\neq a} r^{a c}\Bigg]=}{}
+l \sum\limits_{c\neq a} \sum\limits_{\beta\in R} \sum\limits_{\alpha\in\Pi}\,\sum\limits_{s=0}^{l-1} |\beta|^2
\langle \lambda^s\hat{\alpha},\beta\rangle  f^{m}_{\beta}(z_a - z_c) \p_{u_\alpha} \gt^{m,a}_{\beta}\gt^{-m,c}_{-\beta}.
\end{gather*}

3. Terms $\big[\hat{\p}^a,\frac12f^{bc}\big]$ for $b$, $c\neq a$ and $b\neq c$:
\begin{gather}\label{pr2}
\Bigg[\hat{\p}^a,\frac12\sum\limits_{{b,c\neq a}\atop{b\neq c}}f^{bc}\Bigg]
=\frac{l}{2}\sum\limits_{{b,c\neq a}\atop{b\neq c}}\sum\limits_{k=0}^{l-1}
\sum\limits_{\alpha\in R} |\alpha|^2 \p_{u} f^{k}_{\alpha}(z_b-z_c) \gh^{0,a}_{\alpha}
\gt^{k,b}_{\alpha} \gt^{-k,c}_{-\alpha}.
\end{gather}

3.1. Terms $\big[\hat{\p}^a,\frac12 f^{ac}\big]$ for $c\neq a$:
\begin{gather*}
 \Bigg[\hat{\p}^a,\frac12\sum\limits_{c\neq a}(f^{ac} + f^{ca})\Bigg]
=\Bigg[\hat{\p}^a,\sum\limits_{c\neq a}f^{ac}\Bigg]
=l\sum\limits_{c\neq a}\sum\limits_{m=0}^{l-1}\sum\limits_{\beta\in R} |\beta|^2
\p_{u} f^{m}_{\beta}(z_a - z_c) \gh^{0,a}_{\beta}\gt^{m,a}_{\beta}\gt^{-m,c}_{-\beta}\nonumber\\
 \phantom{\Bigg[\hat{\p}^a,\frac12\sum\limits_{c\neq a}(f^{ac} + f^{ca})\Bigg]}
{}+l\sum\limits_{c\neq a}\sum\limits_{\beta\in R}\sum\limits_{\alpha\in \Pi} \sum\limits_{s=0}^{l-1} |\beta|^2
\langle \lambda^s\hat{\alpha},\beta\rangle f^{m}_{\beta}(z_a - z_c)\p_{u_\alpha} \gt^{m,a}_{\beta}\gt^{-m,c}_{-\beta}.
\end{gather*}
Therefore we get
\begin{gather}\label{pr4}
\Bigg[\hat{\p}^a, \sum\limits_{c\neq a} f^{ac}\Bigg]-\Bigg[\Delta,\sum\limits_{c\neq a}r^{ac}\Bigg]
=\frac{l}{2}\sum\limits_{c\neq a}\sum\limits_{\alpha\in R}\sum\limits_{m=0}^{l-1}
|\alpha|^2\p_{u}f_{\alpha}^{m}(z_a - z_c)
\left[\gh^{0,a}_{\alpha},\gt_{\alpha}^{m,a}\right]_{+} \gt^{-m,c}_{-\alpha}.
\end{gather}

3.2. Terms $\big[\hat{\p}^a,\frac12 f^{aa}\big]$:
\begin{gather}
\left[\hat{\p}^a,\frac12 f^{aa}\right] =-\frac{l}{2}\sum\limits_{m=0}^{l-1}
\sum\limits_{\beta\in R} |\beta|^2
\p_u \wp^{m}_{\beta} \gh^{0,a}_{\beta}\gt^{m,a}_{\beta}\gt^{-m,a}_{-\beta}\nonumber\\
\hphantom{\left[\hat{\p}^a,\frac12 f^{aa}\right]}{}
=-\frac{l}{4}
\sum\limits_{m=0}^{l-1}\sum\limits_{\beta\in R} |\beta|^2
{E_{2}^{\prime}}^{m}_{\beta} \gh^{0,a}_{\beta}
\big[\gt^{m,a}_{\beta}\gt^{-m,a}_{-\beta}\big]_{+}.\label{pr5}
\end{gather}

3.3. Terms $\big[\hat{\p}^a,\frac12f^{cc}\big]$ for $c\neq a$:
\begin{gather}\label{pr52}
\Bigg[\hat{\p}^a, \sum\limits_{c\neq a} \frac12 f^{cc}\Bigg]=
-\frac{l}{4}\sum\limits_{c\neq a}\sum\limits_{m=0}^{l-1}\sum\limits_{\beta\in R} |\beta|^2
{E_{2}^{\prime}}^{m}_{\beta} \gh^{0,a}_{\beta}\big[\gt^{m,c}_{\beta}\gt^{-m,c}_{-\beta}\big]_{+}.
\end{gather}

 4. Terms $[r,f]$:
\begin{gather*}
\Bigg[\sum\limits_{c\neq a}r^{a c},\frac12\sum\limits_{b,d} f^{b,d}\Bigg]=
\frac12 \sum\limits_{{b,c\neq a}\atop {b\neq c}}\big(\big[r^{ac},f^{ab}\big]+
        \big[r^{ac},f^{bc}\big]+\big[r^{ab},f^{ac}\big]+\big[r^{ab},f^{bc}\big]\big)\nonumber\\
\hphantom{\Bigg[\sum\limits_{c\neq a}r^{a c},\frac12\sum\limits_{b,d} f^{b,d}\Bigg]=}{}
      + \sum\limits_{c\neq a}\left(\big[r^{ac},f^{ac}\big]
        +\frac12\big[r^{ac},f^{aa}\big]+\frac12\big[r^{ac},f^{cc}\big]\right).
\end{gather*}

4.1. Terms $\big[r,f\big]$ for $b$, $c\neq a$ and $b\neq c$:
\begin{gather}
\frac12\sum\limits_{{b,c\neq a}\atop {b\neq c}}
\big(\big[r^{ac},f^{ab}\big]+\big[r^{ac},f^{bc}\big]+\big[r^{ab},f^{ac}\big]+\big[r^{ab},f^{bc}\big]\big)\nonumber\\
 \qquad
{}=\frac12\sum\limits_{{b,c\neq a}\atop{b\neq c}}\,\sum\limits_{k,m,s=0}^{l-1}\;\sum\limits_{\alpha,\beta\in R}
|\alpha|^2|\beta|^2\omega^{ks}C_{\lambda^s\alpha,\beta}
\Big(\varphi^{k}_{\alpha}(z_{ac}) f^{m}_{\beta}(z_{ab})-\varphi^{m}_{\beta}(z_{ab}) f^{k}_{\alpha}(z_{ac})\nonumber\\
 \qquad\quad{}
+\varphi^{k+m}_{\alpha+\beta}(z_{ab}) f^{k}_{\alpha}(z_{cb})-\varphi^{k+m}_{\alpha+\beta}(z_{ac})
f^{-m}_{-\beta}(z_{bc})\Big) \gt^{k+m,a}_{\alpha+\lambda^s \beta}
\gt^{-m,b}_{-\beta}\gt^{-k,c}_{-\alpha}\nonumber\\
 \qquad\quad{}
-\frac{l}{2}\sum\limits_{{b,c\neq a}\atop{b\neq c}}\,\sum\limits_{k,m=0}^{l-1} \sum\limits_{\alpha\in R}
|\alpha|^2\Big[\Big( \varphi^{k}_{\alpha}(z_{ac}) f^{m}_{0}(z_{ab})-\varphi^{m}_{0}(z_{ab}) f^{k}_{\alpha}(z_{ac})\nonumber\\
 \qquad\quad{}
+\varphi^{k+m}_{\alpha}(z_{ab}) f^{k}_{\alpha}(z_{cb})
-\varphi^{k+m}_{\alpha}(z_{ac}) f^{-m}_{0}(z_{bc})\Big)
\gt^{k+m,a}_{\alpha} \gh^{-m,b}_{\alpha} \gt^{-k,c}_{\alpha}\nonumber\\
 \qquad\quad{}
+\Big( \varphi^{k}_{\alpha}(z_{ab}) f^{m}_{0}(z_{ac})-\varphi^{m}_{0}(z_{ac}) f^{k}_{\alpha}(z_{ab})\nonumber\\
\qquad\quad{}
+\varphi^{k+m}_{\alpha}(z_{ac}) f^{k}_{\alpha}(z_{bc})-
\varphi^{k+m}_{\alpha}(z_{ac}) f^{-m}_{0}(z_{cb})\Big)\gt^{k+m,a}_{\alpha} \gt^{-k,b}_{\alpha}
\gh^{-m,c}_{\alpha}\nonumber\\
 \qquad\quad{}
+2 \Big( \varphi^{k}_{\alpha}(z_{bc}) f^{m}_{0}(z_{ba}) -\varphi^{m}_{0}(z_{ba}) f^{k}_{\alpha}(z_{bc})\nonumber\\
 \qquad\quad{}
+\varphi^{k+m}_{\alpha}(z_{ba}) f^{k}_{\alpha}(z_{ba})- \varphi^{k+m}_{\alpha}(z_{ba})f^{-m}_{0}(z_{ac})\Big)
\gt^{k+m,b}_{\alpha} \gh^{-m,a}_{\alpha} \gt^{-k,c}_{\alpha}\Big].\label{pr8}
\end{gather}
Almost all terms in this big sum vanish due to the Fay identity (\ref{ap33}), except the terms with $m=0$.
Using
(\ref{ap215}) for these terms in the expression (\ref{pr8}) we get
\begin{gather*}
-\frac{l}{4}\sum\limits_{{b,c\neq a}\atop {b\neq c}}
\sum\limits_{k=0}^{l-1}\sum\limits_{\alpha\in R}|\alpha|^2\p_u f^{k}_{\alpha}(z_{ac})
\gt^{k,a}_{\alpha}\gh^{0,b}_{\alpha}\gt^{-k,c}_{-\alpha}
-\frac{l}{4}\sum\limits_{{b,c\neq a}\atop {b\neq c}}
\sum\limits_{k=0}^{l-1}\sum\limits_{\alpha\in R}|\alpha|^2\p_u f^{k}_{\alpha}(z_{ab})
\gt^{k,a}_{\alpha}\gh^{0,c}_{\alpha}\gt^{-k,b}_{-\alpha}\\
\qquad{}
+\frac{l}{2}\sum\limits_{{b,c\neq a}\atop{b\neq c}}\sum\limits_{k=0}^{l-1}
\sum\limits_{\alpha\in R}|\alpha|^2\p_u f^{k}_{\alpha}(z_{bc})
\gt^{k,b}_{\alpha}\gh^{0,a}_{\alpha}\gt^{-k,c}_{-\alpha}.
\end{gather*}
Finally, using the symmetry in summation over $c$ and $b$ we obtain
\begin{gather}
 \frac12\sum\limits_{{b,c\neq a}\atop {b\neq c}}\big(\big[r^{ac},f^{ab}\big]
+\big[r^{ac},f^{bc}\big]+\big[r^{ab},f^{ac}\big]+\big[r^{ab},f^{bc}\big]\big)\nonumber\\
 \qquad{}
=\frac{l}{2}
\sum\limits_{{b,c\neq a}\atop {b\neq c}}
\sum\limits_{k=0}^{l-1} \sum\limits_{\alpha\in R} |\alpha|^2 \p_u f^{k}_{\alpha}(z_{ac})
\gt^{k,a}_{\alpha} \gh^{0,b}_{\alpha} \gt^{-k,c}_{-\alpha}\nonumber\\
 \qquad\quad{}
-\frac{l}{2}
\sum\limits_{{b,c\neq a}\atop {b\neq c}} \sum\limits_{k=0}^{l-1}
\sum\limits_{\alpha\in R} |\alpha|^2 \p_u f^{k}_{\alpha}(z_{bc})
\gt^{k,b}_{\alpha} \gh^{0,a}_{\alpha} \gt^{-k,c}_{-\alpha}.\label{pr9}
\end{gather}
Notice that the second term here cancels the expression (\ref{pr2}).

4.2. Terms $\big[r^{ac},f^{ac}\big]$:
\begin{gather}
 \sum\limits_{c \neq a}\big[ r^{ac}, f^{ac}\big]
=\frac14\sum\limits_{c \neq a} \sum\limits_{k,m,s=0}^{l-1} \sum\limits_{\alpha,\beta\in R}
|\alpha|^2|\beta|^2\Big(
\varphi^{k}_{\alpha}(z_{ac})f^{m}_{\beta}(z_{ac})\nonumber\\
 \hphantom{\sum\limits_{c \neq a}\big[ r^{ac}, f^{ac}\big]=}{}
-\varphi^{m}_{\beta}(z_{ac})f^{k}_{\alpha}(z_{ac})
\Big)\omega^{ms} C_{\alpha,\lambda^s \beta}\gt^{k+m,a}_{\alpha+\lambda^s\beta}
\big[\gt^{-k,c}_{-\alpha},\gt^{-m,c}_{-\beta}\big]_{+}\nonumber\\
\hphantom{\sum\limits_{c \neq a}\big[ r^{ac}, f^{ac}\big]=}{}
+\frac14\sum\limits_{c \neq a} \sum\limits_{k,m,s=0}^{l-1} \sum\limits_{\alpha,\beta\in R}
|\alpha|^2|\beta|^2\Big(
\varphi^{k}_{\alpha}(z_{ac}) f^{m}_{\beta}(z_{ac})\nonumber\\
\hphantom{\sum\limits_{c \neq a}\big[ r^{ac}, f^{ac}\big]=}{}
-\varphi^{m}_{\beta}(z_{ac})f^{k}_{\alpha}(z_{ac})
\Big)\omega^{-ms} C_{\alpha,\lambda^s \beta}
\gt^{-k-m,c}_{-\alpha-\lambda^s\beta} \big[\gt^{k,a}_{\alpha},\gt^{m,a}_{\beta}\big]_{+}\nonumber\\
\hphantom{\sum\limits_{c \neq a}\big[ r^{ac}, f^{ac}\big]=}{}
-\frac{l}{4}\sum\limits_{c \neq a}  \sum\limits_{k,m=0}^{l-1} \sum\limits_{\beta\in R}
|\beta|^2\Big(\varphi^{-m}_{-\beta}(z_{ac})f^{m+k}_{\beta}(z_{ac})\nonumber\\
\hphantom{\sum\limits_{c \neq a}\big[ r^{ac}, f^{ac}\big]=}{}
-\varphi^{m+k}_{\beta}(z_{ac})f^{-m}_{-\beta}(z_{ac})\Big)\gh^{k,a}_{\beta}
\big[\gt^{-k-m,c}_{-\beta},\gt^{m,c}_{\beta}\big]_{+}\nonumber\\
\hphantom{\sum\limits_{c \neq a}\big[ r^{ac}, f^{ac}\big]=}{}
+\frac{l}{4}\sum\limits_{c \neq a}  \sum\limits_{k,m=0}^{l-1} \sum\limits_{\beta\in R} |\beta|^2
\Big( \varphi^{-m}_{-\beta}(z_{ac})f^{m+k}_{\beta}(z_{ac})\nonumber\\
\hphantom{\sum\limits_{c \neq a}\big[ r^{ac}, f^{ac}\big]=}{}
-\varphi^{m+k}_{\beta}(z_{ac})f^{-m}_{-\beta}(z_{ac})\Big)\gh^{-k,c}_{\beta}
\big[\gt^{k+m,a}_{\beta},\gt^{-m,a}_{-\beta}\big]_{+}\nonumber\\
\hphantom{\sum\limits_{c \neq a}\big[ r^{ac}, f^{ac}\big]=}{}
+\frac{l}{2}\sum\limits_{c \neq a}  \sum\limits_{k,m=0}^{l-1} \sum\limits_{\beta\in R}
|\beta|^2\Big(-\varphi^{-m}_{0}(z_{ac})f^{m+k}_{\beta}(z_{ac})\nonumber\\
\hphantom{\sum\limits_{c \neq a}\big[ r^{ac}, f^{ac}\big]=}{}
+f^{-m}_{0}(z_{ac})\varphi^{m+k}_{\beta}(z_{ac})\Big)
\big[\gh^{-m,a}_{\beta},\gt^{m+k,a}_{\beta}\big]_{+}\gt^{-k,c}_{-\beta}\nonumber\\
\hphantom{\sum\limits_{c \neq a}\big[ r^{ac}, f^{ac}\big]=}{}
+\frac{l}{2}\sum\limits_{c\neq a} \sum\limits_{k,m=0}^{l-1} \sum\limits_{\beta\in R}
|\beta|^2\Big(\varphi^{-m}_{0}(z_{ac})f^{m+k}_{\beta}(z_{ac})\nonumber\\
\hphantom{\sum\limits_{c \neq a}\big[ r^{ac}, f^{ac}\big]=}{}
-f^{-m}_{0}(z_{ac})\varphi^{m+k}_{\beta}(z_{ac})\Big)
\big[\gh^{m,c}_{\beta},\gt^{-m-k,a}_{-\beta}\big]_{+}\gt^{k,c}_{\beta}.\label{pr10}
\end{gather}

4.3. Terms $\big[r^{ac},f^{aa}\big]$ and $\big[r^{ac},f^{cc}\big]$ for $c \neq a$:
\begin{gather}
\Bigg[\sum\limits_{c\neq a} r^{ac},f^{aa}\Bigg]=
 -\frac14\sum\limits_{c\neq a} \sum\limits_{k,m=0}^{l-1}
 \sum\limits_{\alpha,\beta\in R} |\alpha|^2|\beta|^2 \varphi^{k+m}_{\alpha+\beta}(z_{ac})
(\wp^{k}_{\alpha}-\wp^{m}_{\beta})\nonumber\\
\hphantom{\Bigg[\sum\limits_{c\neq a} r^{ac},f^{aa}\Bigg]=}{}
\times\omega^{- m s} C_{\alpha,\lambda^s \beta}
\big[\gt^{k,a}_{\alpha},\gt^{m,a}_{\beta}\big]_{+} \gt^{-k-m,c}_{-\alpha-\lambda^s \beta}\nonumber\\
\hphantom{\Bigg[\sum\limits_{c\neq a} r^{ac},f^{aa}\Bigg]=}{}
-\frac{l}{4} \sum\limits_{c\neq a} \sum\limits_{k,m=0}^{l-1} \sum\limits_{\beta\in R} |\beta|^2
\varphi^{k}_{0}(\wp^{m}_{\beta}-\wp^{k+m}_{\beta})
\gh^{-k,c}_{\beta}\big[\gt^{k+m,a}_{\beta},\gt^{-m,a}_{-\beta}\big]_{+}\nonumber\\
\hphantom{\Bigg[\sum\limits_{c\neq a} r^{ac},f^{aa}\Bigg]=}{}
+\frac{l}{2}\sum\limits_{c\neq a} \sum\limits_{k,m=0}^{l-1} \sum\limits_{\beta\in R}
|\beta|^2\varphi^{k}_{\beta}(z_{ac})(\wp^{m}_{0}-\wp^{m+k}_{\beta})
\big[\gh^{-m,a}_{\beta},\gt^{m+k,a}_{\beta}\big]_{+} \gt^{-k,c}_{-\beta},\label{pr11}
\\
\Bigg[\sum\limits_{c\neq a} r^{ac},f^{cc}\Bigg]=
 -\frac14\sum\limits_{c\neq a} \sum\limits_{k,m=0}^{l-1}
 \sum\limits_{\alpha,\beta\in R} |\alpha|^2|\beta|^2 \varphi^{k+m}_{\alpha+\beta}(z_{ac})
(\wp^{k}_{\alpha}-\wp^{m}_{\beta})\nonumber\\
\hphantom{\Bigg[\sum\limits_{c\neq a} r^{ac},f^{cc}\Bigg]=}{}
\times \omega^{ m s} C_{\alpha,\lambda^s \beta}
\big[\gt^{-k,a}_{-\alpha},\gt^{-m,a}_{-\beta}\big]_{+} \gt^{k+m,c}_{\alpha+\lambda^s \beta}\nonumber\\
\hphantom{\Bigg[\sum\limits_{c\neq a} r^{ac},f^{cc}\Bigg]=}{}
+\frac{l}{4} \sum\limits_{c\neq a} \sum\limits_{k,m=0}^{l-1} \sum\limits_{\beta\in R}
|\beta|^2 \varphi^{k}_{0}(\wp^{m}_{\beta}-\wp^{k+m}_{\beta})
\gh^{k,a}_{\beta}\big[\gt^{-k-m,c}_{-\beta},\gt^{m,c}_{\beta}\big]_{+}\nonumber\\
\hphantom{\Bigg[\sum\limits_{c\neq a} r^{ac},f^{cc}\Bigg]=}{}
-\frac{l}{2} \sum\limits_{c\neq a} \sum\limits_{k,m=0}^{l-1} \sum\limits_{\beta\in R}
|\beta|^2\varphi^{k}_{\beta}(z_{ac})(\wp^{m}_{0}-\wp^{m+k}_{\beta})
\big[\gh^{m,c}_{\beta},\gt^{-m-k,c}_{-\beta}\big]_{+} \gt^{k,a}_{\beta}.\label{pr12}
\end{gather}
The f\/irst two lines in (\ref{pr10}) are canceled by f\/irst lines
in (\ref{pr11}) and (\ref{pr12}) due to iden\-ti\-ty~(\ref{ap29}).
Next, the sum of the third line in (\ref{pr10}), the second
line
in (\ref{pr12}), the sum of fourth line in (\ref{pr10}) and the second line in (\ref{pr11})
are vanished due to (\ref{ap29}) for all values of summation parameters except $k=0$.
For $k=0$ these sums give
\begin{gather*}
 -\frac{l}{4}\sum\limits_{c\neq a} \sum\limits_{m=0}^{l-1} \sum\limits_{\beta\in R}|\beta|^2
\Big(\varphi^{-m}_{-\beta}(z_{ac})f^{m}_{\beta}(z_{ac})-\varphi^{m}_{\beta}(z_{ac})f^{-m}_{-\beta}(z_{ac})\Big)\\
\quad\qquad
{}\times\Big(\gh^{0,a}_{\beta}\big[\gt^{-m,c}_{-\beta},\gt^{m,c}_{\beta}\big]_{+}
-\gh^{0,c}_{\beta}\big[\gt^{m,a}_{\beta},\gt^{-m,a}_{-\beta}\big]_{+}\Big) \\
 \qquad{}
\overset{\eqref{ap210}}{=}
\frac{l}{4}\sum\limits_{c \neq a}  \sum\limits_{m=0}^{l-1} \sum\limits_{\beta\in R}
|\beta|^2{E_{2}^{\prime}}^{m}_{\beta}
\Big(\gh^{0,a}_{\beta}\big[\gt^{-m,c}_{-\beta},\gt^{m,c}_{\beta}\big]_{+}-
\gh^{0,c}_{\beta}\big[\gt^{m,a}_{\beta},\gt^{-m,a}_{-\beta}\big]_{+}\Big)
\end{gather*}
and this is exactly what we need to compensate (\ref{pr5}) and (\ref{pr52}).
(Note that (\ref{pr52}) cancels by the f\/irst
term here and (\ref{pr5}) cancels by the second one due to (\ref{t4}).)

Finally, the last two lines in (\ref{pr10}) are canceled
by the list lines in (\ref{pr11}) and (\ref{pr12}) for all values of
summation parameters except $m=0$.
For $m=0$ the sum of these terms equals
\begin{gather}
 -\frac{l}{2}\sum\limits_{c \neq a}\;\sum\limits_{k=0}^{l-1}\;\sum\limits_{\beta\in R}|\beta|^2
\Big(\varphi^{0}_{0}(z_{ac})f^{k}_{\beta}(z_{ac})-f^{0}_{0}(z_{ac})\varphi^{k}_{\beta}(z_{ac})
+\varphi^{k}_{\beta}(z_{ac})\wp^{k}_{\beta} \Big)\nonumber\\
 \qquad
{}\times\Big(\big[\gh^{0,a}_{\beta},\gt^{k,a}_{\beta}\big]_{+}
\gt^{-k,c}_{-\beta}-\big[\gh^{0,c}_{\beta},\gt^{-k,c}_{-\beta}\big]_{+} \gt^{k,a}_{\beta}\Big).\label{pr13}
\end{gather}
Notice that due to (\ref{t305}) and (\ref{t306}) $f^{0}_{0}(z_{ac})=\rho(z_{ac})$ and
$\varphi^{0}_{0}(z_{ac})=E_1(z_{ac})$.
Then using (\ref{ap211}) we can simplify the expression (\ref{pr13})
\begin{gather}\label{pr14}
-\frac{l}{4}\sum\limits_{c \neq a}
 \sum\limits_{k=0}^{l-1} \sum\limits_{\beta\in R}|\beta|^2 \p_{u}f^{k}_{\beta}(z_{ac})
\Big(\big[\gh^{0,a}_{\beta},\gt^{k,a}_{\beta}\big]_{+}
\gt^{-k,c}_{-\beta}-\big[\gh^{0,c}_{\beta},\gt^{-k,c}_{-\beta}\big]_{+}
\gt^{k,a}_{\beta}\Big).
\end{gather}
Finally, we have nonzero terms from (\ref{pr4}), f\/irst term in (\ref{pr9}) and (\ref{pr14}).
All these terms are proportional
to $\p_{u} f^{k}_{\beta}$. Summing them up we f\/ind
\begin{gather*}
 l\sum\limits_{c \neq a}  \sum\limits_{k=0}^{l-1} \sum\limits_{\beta\in R}
|\beta|^2\p_{u}f^{k}_{\beta}(z_{ac})\Bigg(
{-}\frac14\big[\gh^{0,a}_{\beta},\gt^{k,a}_{\beta}\big]_{+} \gt^{-k,c}_{-\beta}
+\frac14\big[\gh^{0,c}_{\beta},\gt^{-k,c}_{-\beta}\big]_{+} \gt^{k,a}_{\beta}\nonumber\\
 \qquad\quad
{} +\frac12\sum\limits_{b\neq a,c}\gt^{k,a}_{\beta} \gt^{-k,c}_{-\beta} \gh^{0,b}_{\beta}
+\frac12\big[\gh^{0,a}_{\beta},\gt^{k,a}_{\beta}\big]_{+} \gt^{-k,c}_{-\beta}\Bigg)\nonumber\\
 \qquad
{} =l\sum\limits_{c \neq a}  \sum\limits_{k=0}^{l-1} \sum\limits_{\beta\in R}
|\beta|^2 \p_{u}f^{k}_{\beta}(z_{ac})\Bigg(
\frac14\big[\gh^{0,a}_{\beta},\gt^{k,a}_{\beta}\big]_{+} \gt^{-k,c}_{-\beta}\nonumber\\
\qquad\quad{}
+\frac14\big[\gh^{0,c}_{\beta},\gt^{-k,c}_{-\beta}\big]_{+} \gt^{k,a}_{\beta}
+\frac12\sum\limits_{b\neq a,c}\gt^{k,a}_{\beta} \gt^{-k,c}_{-\beta}\gh^{0,b}_{\beta}\Bigg)\\ 
 \qquad
\overset{\eqref{t4}}{=}l\sum\limits_{c \neq a}  \sum\limits_{k=0}^{l-1} \sum\limits_{\beta\in R} |\beta|^2
\p_{u}f^{k}_{\beta}(z_{ac})
\left(\frac14\big[\gh^{0,a}_{\beta},\gt^{k,a}_{\beta}\big]
\gt^{-k,c}_{-\beta}+\frac14\big[\gh^{0,c}_{\beta},\gt^{-k,c}_{-\beta}\big]\gt^{k,a}_{\beta}\right)=0.\tag*{\qed}
\end{gather*}\renewcommand{\qed}{}
\end{proof}

\subsection*{Acknowledgements}

The authors are grateful to A.~Beilinson, L.~Feh\'er, B.~Feigin, A.~Gorsky, S.~Khoroshkin, A.~Losev, A.~Mironov,
V.~Poberezhny, A.~Rosly and A.~Stoyanovsky for useful discussions and remarks.
The work was supported by grants
RFBR-09-02-00393, RFBR-09-01-92437-KE$_a$ and by the Federal Agency for Science and Innovations
of Russian Federation under contract 14.740.11.0347.
The work of A.Z.\ and A.S.\ was also supported by the Russian President fund MK-1646.2011.1,
RFBR-09-01-93106-NCNIL$_a$, RFBR-12-01-00482 and RFBR-12-01-33071 mol\_a\_ved.
The work of A.L.\ was partially supported
by AG Laboratory GU-HSE, RF government grant, ag.\ 11 11.G34.31.0023.

\pdfbookmark[1]{References}{ref}
\LastPageEnding


\begin{thebibliography}{99}
\footnotesize\itemsep=0pt

\bibitem{trig3}
Aminov G., Arthamonov S., Reduction of the elliptic {${\rm SL}(N,{\mathbb C})$}
  top, \href{http://dx.doi.org/10.1088/1751-8113/44/7/075201}{\textit{J.~Phys.~A: Math. Theor.}} \textbf{44} (2011), 075201, 34~pages,
  \href{http://arxiv.org/abs/1009.1867}{arXiv:1009.1867}.

\bibitem{iso24}
Aminov G., Arthamonov S., Levin A.M., Olshanetsky M.A., Zotov A.V., Around
  Painlev{\'e} VI f\/ield theory, submitted.

\bibitem{At}
Atiyah M.F., Vector bundles over an elliptic curve, \href{http://dx.doi.org/10.1112/plms/s3-7.1.414}{\textit{Proc. London Math.
  Soc.}} \textbf{7} (1957), 414--452.

\bibitem{APW}
Axelrod S., Pietra S.D., Witten E., Geometric quantization of Chern--Simons
  gauge theory, \textit{J.~Differential Geom.} \textbf{33} (1991), 787--902.

\bibitem{Bea}
Beauville A., Conformal blocks, fusion rules and the {V}erlinde formula, in
  Proceedings of the {H}irzebruch 65 {C}onference on {A}lgebraic {G}eometry
  ({R}amat {G}an, 1993), \textit{Israel Math. Conf. Proc.}, Vol.~9, Bar-Ilan
  Univ., Ramat Gan, 1996, 75--96, \href{http://arxiv.org/abs/alg-geom/9405001}{alg-geom/9405001}.

\bibitem{BL}
Beauville A., Laszlo Y., Conformal blocks and generalized theta functions,
  \href{http://dx.doi.org/10.1007/BF02101707}{\textit{Comm. Math. Phys.}} \textbf{164} (1994), 385--419,
  \href{http://arxiv.org/abs/alg-geom/9309003}{alg-geom/9309003}.

\bibitem{BF}
Ben-Zvi D., Frenkel E., Geometric realization of the {S}egal--{S}ugawara
  construction, in Topology, geometry and quantum f\/ield theory, \href{http://dx.doi.org/10.1017/CBO9780511526398.006}{\textit{London
  Math. Soc. Lecture Note Ser.}}, Vol.~308, Cambridge University Press,
  Cambridge, 2004, 46--97, \href{http://arxiv.org/abs/math.AG/0301206}{math.AG/0301206}.

\bibitem{Be2}
Bernard D., On the {W}ess--{Z}umino--{W}itten models on {R}iemann surfaces,
  \href{http://dx.doi.org/10.1016/0550-3213(88)90236-2}{\textit{Nuclear Phys.~B}} \textbf{309} (1988), 145--174.

\bibitem{Be1}
Bernard D., On the {W}ess--{Z}umino--{W}itten models on the torus,
  \href{http://dx.doi.org/10.1016/0550-3213(88)90217-9}{\textit{Nuclear Phys.~B}} \textbf{303} (1988), 77--93.

\bibitem{BS1}
Bernstein J., Schwarzman O., Chevalley's theorem for complex crystallographic
  {C}oxeter groups, \href{http://dx.doi.org/10.1007/BF01076385}{\textit{Funct. Anal. Appl.}} \textbf{12} (1978), 308--310.

\bibitem{BS2}
Bernstein J., Schwarzman O., Complex crystallographic {C}oxeter groups and
  af\/f\/ine root systems, \href{http://dx.doi.org/10.2991/jnmp.2006.13.2.2}{\textit{J.~Nonlinear Math. Phys.}} \textbf{13} (2006),
  163--182.

\bibitem{Bou}
Bourbaki N., Lie groups and {L}ie algebras, {C}hapters 4--6, \textit{Elements of
  Mathematics (Berlin)}, Springer-Verlag, Berlin, 2002.

\bibitem{dirac1}
Braden H.W., Dolgushev V.A., Olshanetsky M.A., Zotov A.V., Classical
  {$r$}-matrices and the {F}eigin--{O}desskii algebra via {H}amiltonian and
  {P}oisson reductions, \href{http://dx.doi.org/10.1088/0305-4470/36/25/306}{\textit{J.~Phys.~A: Math. Gen.}} \textbf{36} (2003),
  6979--7000, \href{http://arxiv.org/abs/hep-th/0301121}{hep-th/0301121}.

\bibitem{KW22}
Bulycheva K., Monopole solutions to the Bogomolny equation as three-dimensional
  generalizations of the Kronecker series, \href{http://dx.doi.org/10.1007/s11232-012-0110-x}{\textit{Theoret. and Math. Phys.}}
  \textbf{172} (2012), 1232--1242, \href{http://arxiv.org/abs/1203.4674}{arXiv:1203.4674}.

\bibitem{iso21}
Chernyakov Yu.B., Levin A.M., Olshanetsky M.A., Zotov A.V., Elliptic
  {S}chlesinger system and {P}ainlev\'e~{VI}, \href{http://dx.doi.org/10.1088/0305-4470/39/39/S05}{\textit{J.~Phys.~A: Math. Gen.}}
  \textbf{39} (2006), 12083--12101, \href{http://arxiv.org/abs/nlin.SI/0602043}{nlin.SI/0602043}.

\bibitem{ER}
Enriquez B., Rubtsov V., Hecke--{T}yurin parametrization of the {H}itchin and
  {KZB} systems, in Moscow {S}eminar on {M}athematical {P}hysics. {II},
  \textit{Amer. Math. Soc. Transl. Ser.~2}, Vol.~221, Amer. Math. Soc.,
  Providence, RI, 2007, 1--31, \href{http://arxiv.org/abs/math.AG/9911087}{math.AG/9911087}.

\bibitem{ES1}
Etingof P., Schif\/fmann O., Twisted traces of intertwiners for {K}ac--{M}oody
  algebras and classical dynamical {$R$}-matrices corresponding to generalized
  {B}elavin--{D}rinfeld triples, \textit{Math. Res. Lett.} \textbf{6} (1999),
  593--612, \href{http://arxiv.org/abs/math.QA/9908115}{math.QA/9908115}.

\bibitem{ES2}
Etingof P., Schif\/fmann O., Varchenko A., Traces of intertwiners for quantum
  groups and dif\/ference equations, \href{http://dx.doi.org/10.1023/A:1021619920915}{\textit{Lett. Math. Phys.}} \textbf{62}
  (2002), 143--158, \href{http://arxiv.org/abs/math.QA/0207157}{math.QA/0207157}.

\bibitem{EV}
Etingof P., Varchenko A., Geometry and classif\/ication of solutions of the
  classical dynamical {Y}ang--{B}axter equation, \href{http://dx.doi.org/10.1007/s002200050292}{\textit{Comm. Math. Phys.}}
  \textbf{192} (1998), 77--120, \href{http://arxiv.org/abs/q-alg/9703040}{q-alg/9703040}.

\bibitem{F}
Faltings G., A proof for the {V}erlinde formula, \textit{J.~Algebraic Geom.}
  \textbf{3} (1994), 347--374.

\bibitem{Feher}
Feh{\'e}r L., Pusztai B.G., Generalizations of {F}elder's elliptic dynamical
  {$r$}-matrices associated with twisted loop algebras of self-dual {L}ie
  algebras, \href{http://dx.doi.org/10.1016/S0550-3213(01)00609-5}{\textit{Nuclear Phys.~B}} \textbf{621} (2002), 622--642,
  \href{http://arxiv.org/abs/math.QA/0109132}{math.QA/0109132}.

\bibitem{Fe}
Felder G., The {KZB} equations on {R}iemann surfaces, in Sym\'etries Quantiques
  ({L}es {H}ouches, 1995), North-Holland, Amsterdam, 1998, 687--725,
  \href{http://arxiv.org/abs/hep-th/9609153}{hep-th/9609153}.

\bibitem{FGK}
Felder G., Gawedzki K., Kupiainen A., Spectra of {W}ess--{Z}umino--{W}itten
  models with arbitrary simple groups, \href{http://dx.doi.org/10.1007/BF01228414}{\textit{Comm. Math. Phys.}} \textbf{117}
  (1988), 127--158.

\bibitem{FW}
Felder G., Wieczerkowski C., Conformal blocks on elliptic curves and the
  {K}nizhnik--{Z}amolodchikov--{B}ernard equations, \href{http://dx.doi.org/10.1007/BF02099366}{\textit{Comm. Math. Phys.}}
  \textbf{176} (1996), 133--161, \href{http://arxiv.org/abs/hep-th/9411004}{hep-th/9411004}.

\bibitem{KW12}
Frenkel E., Lectures on the {L}anglands program and conformal f\/ield theory, in
  \href{http://dx.doi.org/10.1007/978-3-540-30308-4_11}{Frontiers in Number Theory, Physics, and Geometry.~{II}}, Springer, Berlin,
  2007, 387--533, \href{http://arxiv.org/abs/hep-th/0512172}{hep-th/0512172}.

\bibitem{FM1}
Friedman R., Morgan J.W., Holomorphic principal bundles over elliptic curves,
  \href{http://arxiv.org/abs/math.AG/9811130}{math.AG/9811130}.

\bibitem{FM2}
Friedman R., Morgan J.W., Holomorphic principal bundles over elliptic
  curves.~{II}. {T}he parabolic construction, \textit{J.~Differential Geom.}
  \textbf{56} (2000), 301--379, \href{http://arxiv.org/abs/math.AG/0006174}{math.AG/0006174}.

\bibitem{FMW}
Friedman R., Morgan J.W., Witten E., Principal {$G$}-bundles over elliptic
  curves, \textit{Math. Res. Lett.} \textbf{5} (1998), 97--118,
  \href{http://arxiv.org/abs/alg-geom/9707004}{alg-geom/9707004}.

\bibitem{Fuchs}
Fuchs J., Schweigert C., The action of outer automorphisms on bundles of chiral
  blocks, \href{http://dx.doi.org/10.1007/s002200050841}{\textit{Comm. Math. Phys.}} \textbf{206} (1999), 691--736,
  \href{http://arxiv.org/abs/hep-th/9805026}{hep-th/9805026}.

\bibitem{Ne11}
Gorsky A., Nekrasov N., Hamiltonian systems of {C}alogero-type, and
  two-dimensional {Y}ang--{M}ills theory, \href{http://dx.doi.org/10.1016/0550-3213(94)90429-4}{\textit{Nuclear Phys.~B}} \textbf{414}
  (1994), 213--238, \href{http://arxiv.org/abs/hep-th/9304047}{hep-th/9304047}.

\bibitem{KW13}
Gukov S., Witten E., Branes and quantization, \textit{Adv. Theor. Math. Phys.}
  \textbf{13} (2009), 1445--1518, \href{http://arxiv.org/abs/0809.0305}{arXiv:0809.0305}.

\bibitem{Ha}
Harnad J., Quantum isomonodromic deformations and the
  {K}nizhnik--{Z}amolodchikov equations, in Sym\-metries and Integrability of
  Dif\/ference Equations ({E}st\'erel, {PQ}, 1994), \textit{CRM Proc. Lecture
  Notes}, Vol.~9, Amer. Math. Soc., Providence, RI, 1996, 155--161,
  \href{http://arxiv.org/abs/hep-th/9406078}{hep-th/9406078}.

\bibitem{Hi}
Hitchin N.J., Flat connections and geometric quantization, \href{http://dx.doi.org/10.1007/BF02161419}{\textit{Comm. Math.
  Phys.}} \textbf{131} (1990), 347--380.

\bibitem{Hi1}
Hitchin N.J., Stable bundles and integrable systems, \href{http://dx.doi.org/10.1215/S0012-7094-87-05408-1}{\textit{Duke Math.~J.}}
  \textbf{54} (1987), 91--114.


\bibitem{Ho}
Hori K., Global aspects of gauged {W}ess--{Z}umino--{W}itten models,
  \href{http://dx.doi.org/10.1007/BF02506384}{\textit{Comm. Math. Phys.}} \textbf{182} (1996), 1--32,
  \href{http://arxiv.org/abs/hep-th/9411134}{hep-th/9411134}.

\bibitem{I}
Ivanov D.A., Knizhnik--{Z}amolodchikov--{B}ernard equations on {R}iemann
  surfaces, \href{http://dx.doi.org/10.1142/S0217751X95001200}{\textit{Internat.~J. Modern Phys.~A}} \textbf{10} (1995),
  2507--2536, \href{http://arxiv.org/abs/hep-th/9410091}{hep-th/9410091}.

\bibitem{KW23}
Ivanova T.A., Lechtenfeld O., Popov A.D., Rahn T., Instantons and
  {Y}ang--{M}ills f\/lows on coset spaces, \href{http://dx.doi.org/10.1007/s11005-009-0336-1}{\textit{Lett. Math. Phys.}} \textbf{89}
  (2009), 231--247, \href{http://arxiv.org/abs/0904.0654}{arXiv:0904.0654}.

\bibitem{Ka}
Kac V.G., Inf\/inite-dimensional {L}ie algebras, 3rd ed., \href{http://dx.doi.org/10.1017/CBO9780511626234}{Cambridge University
  Press}, Cambridge, 1990.

\bibitem{KW11}
Kapustin A., Witten E., Electric-magnetic duality and the geometric {L}anglands
  program, \textit{Commun. Number Theory Phys.} \textbf{1} (2007), 1--236,
  \href{http://arxiv.org/abs/hep-th/0604151}{hep-th/0604151}.

\bibitem{KZ}
Knizhnik V.G., Zamolodchikov A.B., Current algebra and {W}ess--{Z}umino model
  in two dimensions, \href{http://dx.doi.org/10.1016/0550-3213(84)90374-2}{\textit{Nuclear Phys.~B}} \textbf{247} (1984), 83--103.

\bibitem{Ta1}
Korotkin D., Samtleben H., On the quantization of isomonodromic deformations on
  the torus, \href{http://dx.doi.org/10.1142/S0217751X97001274}{\textit{Internat.~J. Modern Phys.~A}} \textbf{12} (1997),
  2013--2029, \href{http://arxiv.org/abs/hep-th/9511087}{hep-th/9511087}.

\bibitem{KT}
Kuroki G., Takebe T., Twisted {W}ess--{Z}umino--{W}itten models on elliptic
  curves, \href{http://dx.doi.org/10.1007/s002200050233}{\textit{Comm. Math. Phys.}} \textbf{190} (1997), 1--56,
  \href{http://arxiv.org/abs/q-alg/9612033}{q-alg/9612033}.

\bibitem{Ne13}
Levin A.M., Olshanetsky M.A., Double coset construction of moduli space of
  holomorphic bundles and {H}itchin systems, \href{http://dx.doi.org/10.1007/s002200050173}{\textit{Comm. Math. Phys.}}
  \textbf{188} (1997), 449--466, \href{http://arxiv.org/abs/alg-geom/9605005}{alg-geom/9605005}.

\bibitem{LO}
Levin A.M., Olshanetsky M.A., Hierarchies of isomonodromic deformations and
  {H}itchin systems, in Moscow {S}eminar in {M}athematical {P}hysics,
  \textit{Amer. Math. Soc. Transl. Ser.~2}, Vol.~191, Amer. Math. Soc.,
  Providence, RI, 1999, 223--262.

\bibitem{LOSZ2}
Levin A.M., Olshanetsky M.A., Smirnov A.V., Zotov A.V., Calogero--{M}oser
  systems for simple {L}ie groups and characteristic classes of bundles,
  \href{http://dx.doi.org/10.1016/j.geomphys.2012.03.012}{\textit{J.~Geom. Phys.}} \textbf{62} (2012), 1810--1850, \href{http://arxiv.org/abs/1007.4127}{arXiv:1007.4127}.

\bibitem{LOSZ}
Levin A.M., Olshanetsky M.A., Smirnov A.V., Zotov A.V., Characteristic classes
  and {H}itchin systems. {G}eneral construction, \href{http://dx.doi.org/10.1007/s00220-012-1585-x}{\textit{Comm. Math. Phys.}}
  \textbf{316} (2012), 1--44, \href{http://arxiv.org/abs/1006.0702}{arXiv:1006.0702}.

\bibitem{int2}
Levin A.M., Olshanetsky M.A., Smirnov A.V., Zotov A.V., Characteristic classes
  of ${\rm SL}(N)$-bundles and quantum dynamical elliptic $R$-matrices,
  \textit{J.~Phys.~A: Math. Theor.}, {t}o appear, \href{http://arxiv.org/abs/1208.5750}{arXiv:1208.5750}.


\bibitem{LOZ_1}
Levin A.M., Olshanetsky M.A., Zotov A.V., Hitchin systems~-- symplectic {H}ecke
  correspondence and two-dimensional version, \href{http://dx.doi.org/10.1007/s00220-003-0801-0}{\textit{Comm. Math. Phys.}}
  \textbf{236} (2003), 93--133, \href{http://arxiv.org/abs/nlin.SI/0110045}{nlin.SI/0110045}.

\bibitem{KW21}
Levin A.M., Olshanetsky M.A., Zotov A.V., Monopoles and modif\/ications of
  bundles over elliptic curves, \href{http://dx.doi.org/10.3842/SIGMA.2009.065}{\textit{SIGMA}} \textbf{5} (2009), 065,
  22~pages, \href{http://arxiv.org/abs/0811.3056}{arXiv:0811.3056}.

\bibitem{iso12}
Levin A.M., Olshanetsky M.A., Zotov A.V., Painlev\'e {VI}, rigid tops and
  ref\/lection equation, \href{http://dx.doi.org/10.1007/s00220-006-0089-y}{\textit{Comm. Math. Phys.}} \textbf{268} (2006), 67--103,
  \href{http://arxiv.org/abs/math.QA/0508058}{math.QA/0508058}.


\bibitem{iso22}
Levin A.M., Zotov A.V., On rational and elliptic forms of {P}ainlev\'e {VI}
  equation, in Moscow {S}eminar on {M}athematical {P}hysics.~{II},
  \textit{Amer. Math. Soc. Transl. Ser.~2}, Vol.~221, Amer. Math. Soc.,
  Providence, RI, 2007, 173--183.

\bibitem{Loo}
Looijenga E., Root systems and elliptic curves, \href{http://dx.doi.org/10.1007/BF01390167}{\textit{Invent. Math.}}
  \textbf{38} (1976), 17--32.

\bibitem{agt3}
Mironov A., Morozov A., Runov B., Zenkevich Y., Zotov A., Spectral duality
  between Heisenberg chain and Gaudin model, \href{http://dx.doi.org/10.1007/s11005-012-0595-0}{\textit{Lett. Math. Phys.}}, {t}o
  appear, \href{http://arxiv.org/abs/1206.6349}{arXiv:1206.6349}.

\bibitem{agt1}
Mironov A., Morosov A., Shakirov Sh., Towards a proof of {AGT} conjecture by
  methods of matrix models, \href{http://dx.doi.org/10.1142/S0217751X12300013}{\textit{Internat.~J. Modern Phys.~A}} \textbf{27}
  (2012), 1230001, 32~pages, \href{http://arxiv.org/abs/1011.5629}{arXiv:1011.5629}.


\bibitem{agt2}
Mironov A., Morozov A., Zenkevich Y., Zotov A., Spectral duality in integrable
  systems from AGT conjecture, \textit{JETP Lett.}, to appear, \href{http://arxiv.org/abs/1204.0913}{arXiv:1204.0913}.

\bibitem{NS}
Narasimhan M.S., Seshadri C.S., Stable and unitary vector bundles on a compact
  {R}iemann surface, \href{http://dx.doi.org/10.2307/1970710}{\textit{Ann. of Math.~(2)}} \textbf{82} (1965), 540--567.

\bibitem{Ne12}
Nekrasov N., Holomorphic bundles and many-body systems, \href{http://dx.doi.org/10.1007/BF02099624}{\textit{Comm. Math.
  Phys.}} \textbf{180} (1996), 587--603, \href{http://arxiv.org/abs/hep-th/9503157}{hep-th/9503157}.

\bibitem{KW14}
Nekrasov N., Pestun V., Seiberg--Witten geometry of four dimensional $N=2$
  quiver gauge theories, \href{http://arxiv.org/abs/1211.2240}{arXiv:1211.2240}.

\bibitem{int12}
Olshanetsky M.A., Three lectures on classical integrable systems and gauge
  f\/ield theories, \href{http://dx.doi.org/10.1134/S1063779609010067}{\textit{Phys. Part. Nuclei}} \textbf{40} (2009), 93--114,
  \href{http://arxiv.org/abs/0802.3857}{arXiv:0802.3857}.

\bibitem{iso23}
Olshanetsky M.A., Zotov A. V., Isomonodromic problems on elliptic curve, rigid
  tops and ref\/lection equations, in Elliptic Integrable Systems, \textit{Rokko Lectures in Mathematics}, Vol.~18, Kobe University, 2005, 149--172.

\bibitem{PS}
Presley A., Segal G., Loop groups, Clarendon Press, Oxford, 1986.

\bibitem{Re}
Reshetikhin N., The {K}nizhnik--{Z}amolodchikov system as a deformation of the
  isomonodromy problem, \href{http://dx.doi.org/10.1007/BF00420750}{\textit{Lett. Math. Phys.}} \textbf{26} (1992),
  167--177.

\bibitem{Sch}
Schweigert C., On moduli spaces of f\/lat connections with non-simply connected
  structure group, \href{http://dx.doi.org/10.1016/S0550-3213(97)00152-1}{\textit{Nuclear Phys.~B}} \textbf{492} (1997), 743--755,
  \href{http://arxiv.org/abs/hep-th/9611092}{hep-th/9611092}.

\bibitem{Si}
Simpson C.T., Harmonic bundles on noncompact curves, \href{http://dx.doi.org/10.2307/1990935}{\textit{J.~Amer. Math.
  Soc.}} \textbf{3} (1990), 713--770.

\bibitem{trig2}
Smirnov A.V., Integrable {${\rm sl}(N,{\mathbb C})$}-tops as
  {C}alogero--{M}oser systems, \href{http://dx.doi.org/10.1007/s11232-009-0024-4}{\textit{Theoret. and Math. Phys.}} \textbf{158}
  (2009), 300--312, \href{http://arxiv.org/abs/0809.2187}{arXiv:0809.2187}.

\bibitem{Ta2}
Takasaki K., Gaudin model, {KZ} equation and an isomonodromic problem on the
  torus, \href{http://dx.doi.org/10.1023/A:1007417518021}{\textit{Lett. Math. Phys.}} \textbf{44} (1998), 143--156,
  \href{http://arxiv.org/abs/hep-th/9711058}{hep-th/9711058}.

\bibitem{iso41}
Zabrodin A.V., Zotov A.V., Quantum Painlev\'e--Calogero correspondence,
  \href{http://dx.doi.org/10.1063/1.4732532}{\textit{J.~Math. Phys.}} \textbf{53} (2012), 073507, 19~pages,
  \href{http://arxiv.org/abs/1107.5672}{arXiv:1107.5672}.

\bibitem{iso42}
Zabrodin A.V., Zotov A.V., Quantum Painlev\'e--Calogero correspondence for
  Painlev\'e~VI, \href{http://dx.doi.org/10.1063/1.4732534}{\textit{J.~Math. Phys.}} \textbf{53} (2012), 073508, 19~pages,
  \href{http://arxiv.org/abs/1107.5672}{arXiv:1107.5672}.

\bibitem{LOZ_3}
Zotov A.V., {$1+1$} {G}audin model, \href{http://dx.doi.org/10.3842/SIGMA.2011.067}{\textit{SIGMA}} \textbf{7} (2011), 067,
  26~pages, \href{http://arxiv.org/abs/1012.1072}{arXiv:1012.1072}.

\bibitem{int13}
Zotov A.V., Classical integrable systems and their f\/ield-theoretical
  generalizations, \href{http://dx.doi.org/10.1134/S1063779606030063}{\textit{Phys. Part. Nuclei}} \textbf{37} (2006), 759--842.

\bibitem{iso11}
Zotov A.V., Elliptic linear problem for the {C}alogero--{I}nozemtsev model and
  {P}ainlev\'e {VI} equation, \href{http://dx.doi.org/10.1023/B:MATH.0000032753.97756.94}{\textit{Lett. Math. Phys.}} \textbf{67} (2004),
  153--165, \href{http://arxiv.org/abs/hep-th/0310260}{hep-th/0310260}.

\bibitem{trig1}
Zotov A.V., Chernyakov Yu.B., Integrable multiparticle systems obtained by means
  of the {I}nozemtsev limit, \href{http://dx.doi.org/10.1023/A:1012835207484}{\textit{Theoret. and Math. Phys.}} \textbf{129}
  (2001), 1526--1542, \href{http://arxiv.org/abs/hep-th/0102069}{hep-th/0102069}.

\bibitem{int11}
Zotov A.V., Levin A.M., An integrable system of interacting elliptic tops,
  \href{http://dx.doi.org/10.1007/s11232-006-0005-9}{\textit{Theoret. and Math. Phys.}} \textbf{146} (2006), 45--52.

\bibitem{dirac2}
Zotov A.V., Levin A.M., Olshanetsky M.A., Chernyakov Yu.B., Quadratic algebras
  associated with elliptic curves, \href{http://dx.doi.org/10.1007/s11232-008-0081-0}{\textit{Theoret. and Math. Phys.}}
  \textbf{156} (2008), 1103--1122, \href{http://arxiv.org/abs/0710.1072}{arXiv:0710.1072}.

\end{thebibliography}
\end{document}